\documentclass[12pt,preprint]{aastex}
\usepackage{float,epsfig}
\usepackage{graphicx}
\usepackage{color}
\usepackage{natbib}
\usepackage[normalem]{ulem}

\usepackage{graphicx}
\usepackage{subfigure}
\usepackage{amssymb}
\usepackage{amsmath}
\usepackage{mathrsfs}
\usepackage{multirow}
\usepackage{array}
\usepackage{threeparttable}
\usepackage{color}
\usepackage{hyperref}
\usepackage{lineno}

\newcommand{\degree}{^{\circ}}

\slugcomment{\color{blue} Version 8: 2015.04.17}

\begin{document}


\title{Characterization of the Inner Knot of the Crab: The Site of the Gamma-ray Flares?}


\author{
Alexander Rudy\altaffilmark{1},
Dieter Horns\altaffilmark{2},
Andrea DeLuca\altaffilmark{3},
Jeffery Kolodziejczak\altaffilmark{4},
Allyn Tennant\altaffilmark{4},
Yajie Yuan\altaffilmark{7},
Rolf Buehler\altaffilmark{5},
Jonathon Arons\altaffilmark{6},
Roger Blandford\altaffilmark{7},
Patrizia Caraveo\altaffilmark{3},
\\Enrico Costa\altaffilmark{8},
Stephan Funk\altaffilmark{7},
Elizabeth Hays\altaffilmark{11},
Andrei Lobanov\altaffilmark{7},
Claire Max\altaffilmark{1},
\\Michael Mayer\altaffilmark{5},
Roberto Mignani\altaffilmark{10},
Stephen L. O'Dell\altaffilmark{4},
Roger Romani\altaffilmark{7},
Marco Tavani\altaffilmark{3},
\\ Martin C. Weisskopf\altaffilmark{4}
}

\altaffiltext{1}
{Department of Astronomy \& Astrophysics, University of California, Santa Cruz, CA 95064, USA}
\altaffiltext{2}
{Institut f\"{u}r Experimentalphysik, Universit\"{a}t Hamburg, Luruper Chaussee 149, D-22761 Hamburg, Germany}
\altaffiltext{3}
{INAF--IASF Milano, via E. Bassini 15, I-20133 Milano, Italy; INFN Pavia, via A. Bassi 6, I-27100 Pavia, Italy}
\altaffiltext{4}
{NASA Marshall Space Flight Center, Astrophysics Office (ZP12), Huntsville, AL 35812, USA}
\altaffiltext{5}
{DESY, Platanenallee 6, D-15738 Zeuthen, Germany}
\altaffiltext{6}
{Astronomy Department and Theoretical Astrophysics Center, University of California, Berkeley, 601 Campbell Hall, Berkeley, CA 94720, USA}
\altaffiltext{7}
{W. W. Hansen Experimental Physics Laboratory, Kavli Institute for Particle Astrophysics and Cosmology, Stanford University, Stanford, CA 94305, USA}
\altaffiltext{8}
{INFN Roma Tor Vergata, via della Ricerca Scientifica 1, I-00133 Roma, Italy}
\altaffiltext{9}
{Max-Planck-Institut f\"{u}r Radioastronomie, auf dem H\"{u}gel 69, D-53121 Bonn, Germany}
\altaffiltext{10}
{INAF--IASF Milano, via E. Bassini 15, I-20133 Milano, Italy; Kepler Institute of Astronomy, University of Zielona G´ora, Lubuska 2, 65-265 Zielona G´ora, Poland}
\altaffiltext{11}
{NASA Goddard Space Flight Center, Astrophysics Science Division, Greenbelt, MD 20771, USA}




\begin{abstract}

One of the most intriguing results from the $\gamma$-ray instruments currently in orbit has been the detection of powerful flares from the Crab Nebula. 
Such events, with a cadence of about one per year, can be so dramatic as to make the system the brightest source in the $\gamma$-ray sky as occurred in 2011 April.
These flares challenge our understanding of pulsar wind nebulae and models for particle acceleration. 
To pinpoint the production site(s) within the Nebula, a multiwavelength campaign has been carried out using Keck, {\sl Hubble Space Telescope} (HST), and {\sl Chandra X-ray Observatory}. 
As the short time scales of the flares ($\lesssim1$ day) suggest a small emitting region, the Crab's inner knot, located within a fraction of an arcsecond from the pulsar, has been proposed to be just such a site. In this paper, we focus on IR, optical, and X-ray observations of this feature to see if it might be related to the $\gamma$-ray flares and to try to understand its nature. 
We find that the knot's radial size, tangential size, peak flux, and the ratio of the knot flux to that of the pulsar as measured with HST are all correlated with the projected distance of the knot from the pulsar. 
A new approach, using singular value decomposition for analyzing time series of images, was introduced yielding results consistent with the more traditional methods while some uncertainties were substantially
reduced.

We exploit the new characterization of the Crab's inner knot to discuss constraints on standard shock-model parameters that may be inferred from our observations assuming the inner knot lies near to the shocked surface. 
These include inferences as to wind magnetization $\sigma$, shock shape parameters such as incident angle $\delta_1$ and poloidal radius of curvature $R_c$, as well as the IR/optical emitting particle enthalpy fraction. 
We find that while the standard shock model gives good agreement with observation in many respects, there remain two puzzles: 
(a) the observed angular size of the knot relative to the pulsar--knot separation is measured to be much smaller than expected; 
(b) the variable, yet high degree of polarization reported is difficult to reconcile with a highly relativistic downstream flow.
In contrast, the IR/optical flux of the inner knot is marginally consistent with the scenario that the shock accelerates most of the optical emitting particles in the nebula.

\end{abstract}





\section{Introduction}\label{s:intro}

The Crab pulsar and its nebula are one of the most studied targets in the sky at all wavelengths. 
It serves as a test bed for pulsar theories as well as, more generally, for astrophysical non-thermal processes. 
A very thorough review of the Crab system, including high spatial resolution observations with HST and {\sl Chandra}, showing the complex, dynamic interaction of the pulsar wind with the surrounding medium, has been compiled by \citet{Hest08}. 
A more recent review by \citet{Bueh14} also includes the saga of the $\gamma$-ray flares. 

A crucial question implied by the observations of the $\gamma$-ray flares is ``how are the particles being accelerated to sustain the flaring behavior?''
Radiating particles must have PeV energies to emit synchrotron radiation at the observed $\gamma$-ray wavelengths.
A mechanism is needed that can accelerate particles to such high energies in less than one gyration time in the ambient magnetic field. 
For example, the large flare of 2011 April produced a radiant energy equivalent to the energy stored in the mean magnetic field within a region of order $2\times10^{14}$ m subtending an angle $\approx0.3\arcsec$ at 2~kpc, consistent with the emission-region size estimated from light-travel-time arguments \citep{Weis13}.

An intense theoretical effort is currently being devoted to explain the variable Crab behavior \citep[e.g.,][]{Uzdensky2011ApJ...737L..40U, Yuan2011ApJ...730L..15Y, KomissarovLyutikov2011MNRAS.414.2017K, Arons2012SSRv..173..341A, Sturrock2012ApJ...751L..32S, Lyubarsky2012MNRAS.427.1497L, Lyutikov2012MNRAS.422.3118L, Bykov2012MNRAS.421L..67B, Clausen-Brown2012MNRAS.426.1374C, Cerutti2012ApJ...746..148C, Cerutti2013ApJ...770..147C, Baty2013MNRAS.436L..20B, Teraki2013ApJ...763..131T}.
However, the mechanism driving the flares, their impulsive nature, the $\approx 12$-month recurrence time, and the location, remain unknown.
Possible explanations include plasma instabilities in the nebula, magnetic reconnection, discontinuity in the pulsar wind acceleration, emission from the anvil, emission from the inner knot, or portions of the termination shock. 
See, for example, the review by \citet{Hest08} for the association of specific features with this nomenclature

There are a number of reasons to suspect that the feature known as the "inner knot" should be considered as a possible site for the origin of the flares \citep[e.g.,][]{ KomissarovLyutikov2011MNRAS.414.2017K}.
First, the luminosity of the flares accounts for a relatively high fraction of the pulsar spin-down luminosity (assuming a radiation efficiency of 3--5\%), suggesting an origin in a region close to the pulsar \citep{Tava11}. 
Second, the very short variability time scales combined with causality arguments, strongly constrain the size of the flare's emitting region \citep{Bueh12,Stri11}.

A multiwavelength campaign to study the Crab, described in \citet{Weis13}, is being performed. 
Another $\gamma$-ray flare---the second largest observed to date---was detected with {\sl Fermi}/LAT in 2013 March \citep{Maye13}. 
This triggered further observations with Keck, with the {\sl Hubble Space Telescope} (HST), and with the {\sl Chandra X-ray Observatory}. 
Here we concentrate on what was learned about the inner knot as a consequence of this campaign.
We note that very-high-energy ($E \geq 100$~GeV) contemporaneous data from the HESS \citep{Abra14} and VERITAS \citep{Aliu14} Cherenkov-telescope arrays found no significant correlation of TeV flux with the flare observed with {\sl Fermi}/LAT.

The observations are presented in \S\ref{s:obs}. 
The measured properties of the knot are given in \S\ref{s:prop}.
\S\ref{s:compkhc} compares observations of the inner knot at the different wavelengths.
Comparison and examination of possible correlations of the knot's properties with $\gamma$-rays detected using the {\sl Fermi}/LAT are discussed in \S\ref{s:comp}.
\S\ref{s:implications} discusses theoretical implications.
\S\ref{s:conclusions} summarizes.

\section{The Observations} \label{s:obs}

The feature we refer to as the ``inner knot'' (aka ``knot 1'', or ``synchrotron knot'') was discovered by \citet{Hest95} in a detailed study of the Crab using high-resolution images taken with HST. 
The feature, although resolved in those data, is small ($\approx 0.1\arcsec$) and close to the pulsar ($\approx 0.6\arcsec$). 
The feature was first suggested to be associated with a shock or instability in the jet $\approx1500$ AU from the pulsar, due to its good alignment with the jet and elongation perpendicular to the jet. 
Later \citet{Komi03} proposed, based on two-dimensional magnetohydrodynamic (MHD) simulation of the pulsar wind nebula, that the knot represents Doppler-boosted emission from an oblique termination shock. 
Synthetic synchrotron images from these 2D simulations \citep{KomissarovLyutikov2011MNRAS.414.2017K} and more recently 3D simulations \citep{Porth14} indeed show a knot-like feature when the system is viewed from the proper (and reasonable) orientation.  
Moreover, the observed high degree of polarization and a position angle aligned with the symmetry axis \citep{Moran2013MNRAS.433.2564M} seems to lend further support to the oblique-shock scenario.

The salient features of the inner knot deduced prior to our observations are that it has a power-law spectrum $F_{\nu}\propto\nu^{-\alpha}$ with index $\alpha\sim0.8$ in the IR to optical (\citet{Soll03}, \citet{Melatos05}) 
and, as most recently measured by \citet{Moran2013MNRAS.433.2564M}, the flux varies, the position appears to change, the degree of polarization is very large but does not appear to change with the flux (but see the discussion at the end of \S\ref{s:comp}), and the magnetic field (determined from the position angle of the optical polarization) is orthogonal to the axis of symmetry (presumably the spin axis) and consistent with the direction of the field for the rest of the inner nebula. 

\subsection{Keck}\label{ss:keckobs}

We obtained K' and H band near-infrared (NIR) images of the inner $40\arcsec$ of the Crab Nebula using the Keck Near-Infrared Camera 2 (NIRC2; K. Matthews, PI), with laser-guide-star adaptive optics \citep[AO;][]{2006PASP..118..297W}. 
From 2012 February to 2014 January, we obtained 11 observations of the nebula during $\gamma$-ray quiescent periods and 1 target-of-opportunity observation (K8) triggered by $\gamma$-ray flares. 
Table~\ref{t:keckobs} summarizes the Keck observations.

Data reduction was carried out using the pipeline developed by \citet{Ghez08}. 
Individual frames were sky-subtracted, flat-fielded, bad-pixel-corrected, and corrected for the NIRC2's camera distortion.
We used the solution developed by \citet{Yeld10}, which matched the positions of stars in the globular cluster M92 measured with HST to those measured in the NIRC2. 
Frames were checked to ensure that they showed no irregularities in the point-spread function (PSF), and that the Strehl ratio was at least $\sim0.10$.
Individual frames were then combined using the Drizzle algorithm \citep{Fruc02} to provide sub-pixel accuracy and high signal-to-noise images for analysis and modeling (\S\ref{ss:keckprop} and Appendix). 

The observations were conducted with two different pixel scales, using the NIRC2 ``wide'' camera, with $0.04\arcsec$ pixels and a $40\arcsec\times40\arcsec$ field of view, and the NIRC2 ``narrow'' camera, with $0.01\arcsec$ pixels and a $10\arcsec\times10\arcsec$ field of view. 
The total effective exposure times ranged from $420$ to $\approx2600$ seconds.
The shorter exposures result in optimal use of 1-hour Target-of-Opportunity interrupts---including acquisition, calibration, and sky measurements. 

\begin{figure}[ht]
\begin{center}
\includegraphics[angle=0,width=1\columnwidth]{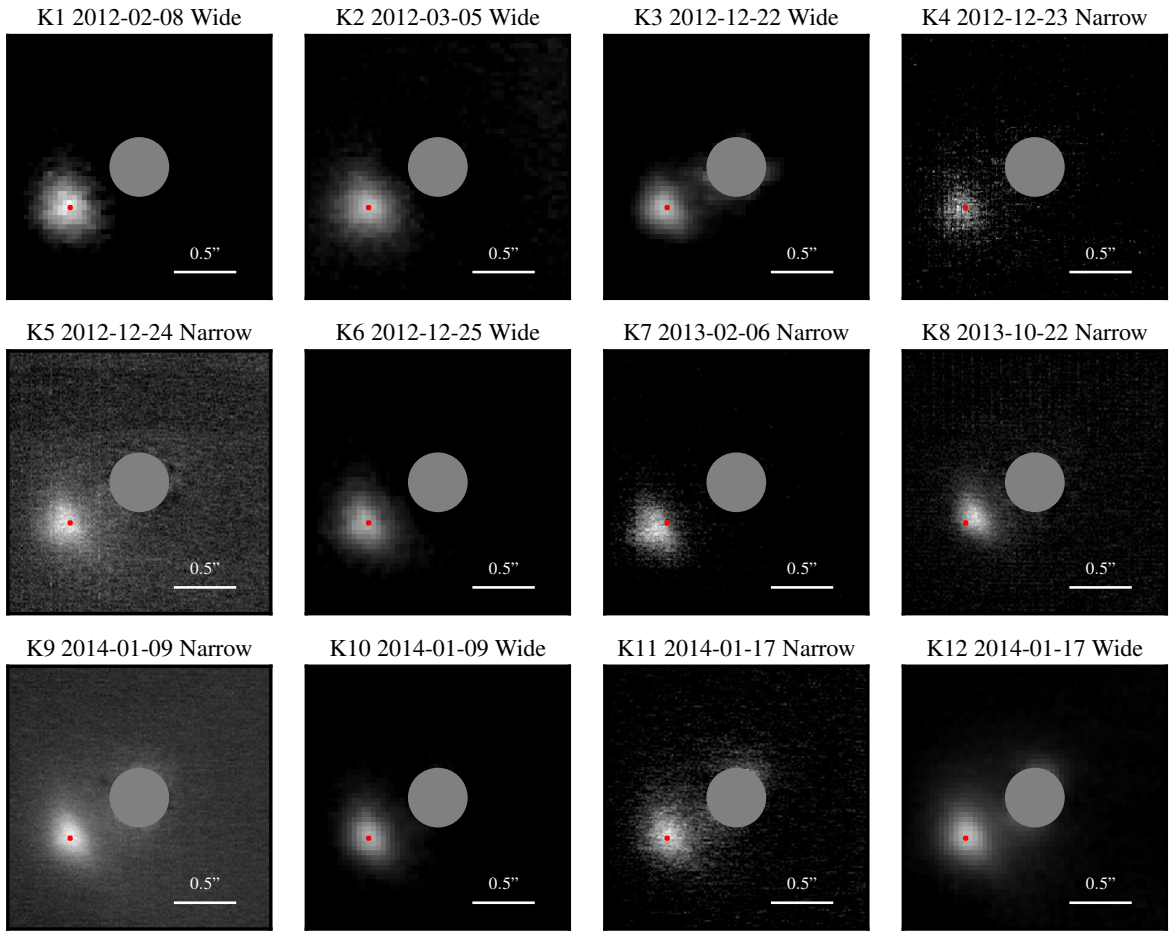}
\figcaption{Individual images of the inner knot, with the pulsar removed, from Keck NIRC2 with laser-guide-star adaptive optics (AO). 
In each image, the PSF of the pulsar has been subtracted using the scaled PSF of a nearby star and replaced with a gray circle at the original location of the pulsar. 
Each image is $2.2\arcsec$ by $2.2\arcsec$. 
Images are from the NIRC2 wide camera ($0.04\arcsec$ pixels) or the NIRC2 narrow camera ($0.01\arcsec$ pixels). 
All images were taken using the $K^\prime$ filter. 
A red dot marks the average position of the inner knot across all 12 observations.
\label{f:keck1}}
\end{center}
\end{figure}

The spatial resolution of Keck's NIRC2 with AO proved to be very effective in resolving the detailed structure and position of the knot. 
Figure~\ref{f:keck1} shows the 12 NIRC2 images.  
In the narrow camera, the center of the pulsar and the center of the knot are separated by $\approx60$ pixels. 
Typical values for the signal-to-noise ratio of the knot are $5$. 
The PSF of NIRC2 with AO varies slightly in time due to the variable performance of the AO system. 
As the pulsar and knot are well resolved from each other in the Keck images, we chose not to deconvolve the PSF from the Keck images.

\subsection{HST}\label{ss:hstobs}

Since 2012 January, we have obtained 17 observations using the HST Advanced Camera for Surveys (ACS) Wide-Field Channel (WFC), with $0.05\arcsec$ pixels. 
Each observation comprises 4 exposures totaling 2000 seconds and uses a standard 4-point BOX dithering pattern to fill the inter-chip gap as well as to allow efficient cosmic-ray cleaning. 
We employed the F550M filter, which is well suited to sampling continuum emission from the Crab with almost no contamination from line emission. 

\begin{figure}[ht]
\begin{center}
\includegraphics[angle=0,width=1.0\columnwidth]{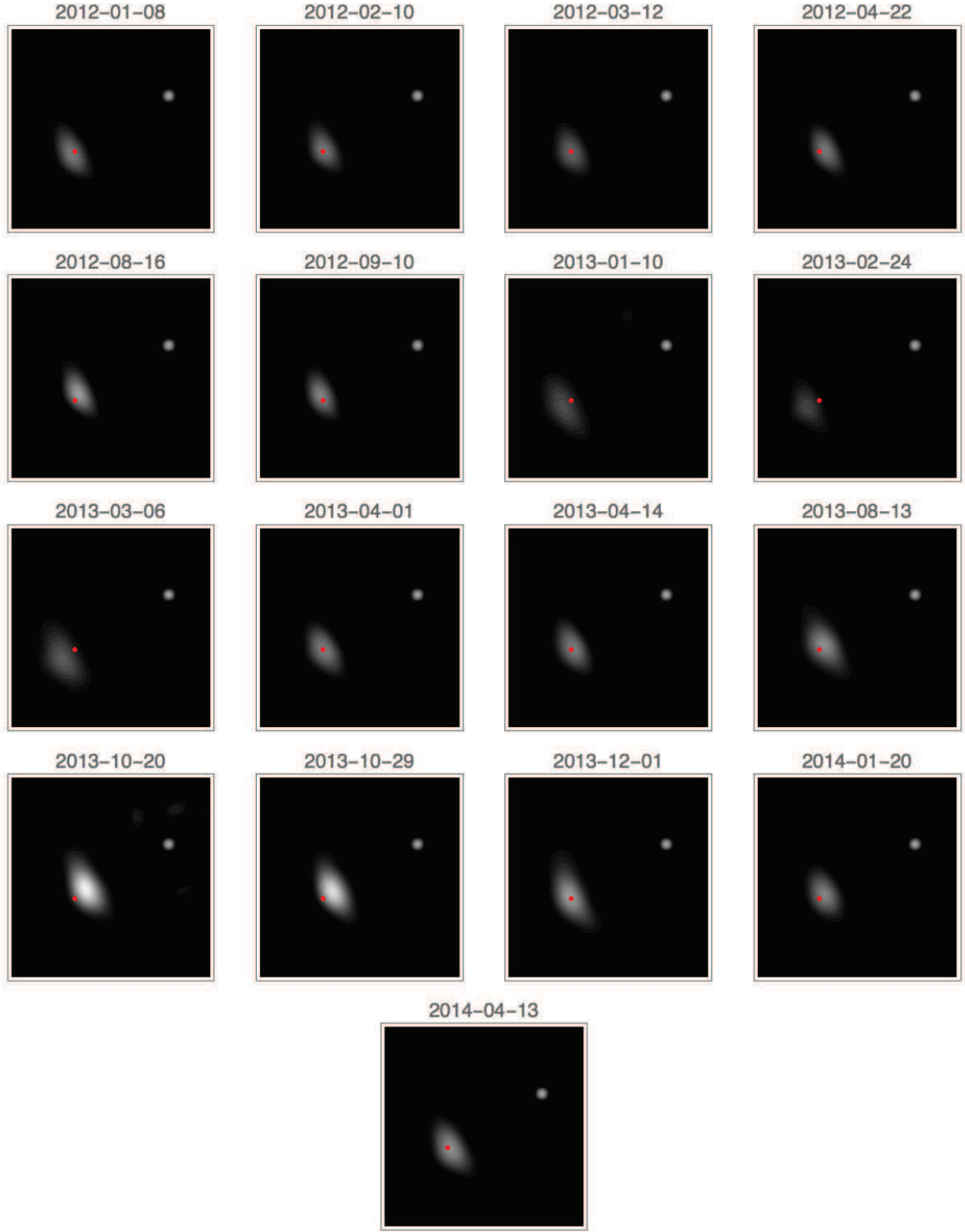}
\figcaption{HST-SVD-Processed images of the Crab pulsar and inner knot after removing effects of the HST/ACS WFC point spread function.
 A red dot marks the average position of the inner knot across all 17 observations.
\label{f:hst1}}
\end{center}
\end{figure}

Table~\ref{t:hstobs} summarizes the observations, listing the date and the mean observing time.
The Appendix  discusses in detail our analyses of the 17 HST/ACS images. 
Figure~\ref{f:hst1} displays the processed images, which have removed effects of the HST/ACS WFC point spread function using singular-value decomposition (SVD) and Richardson-Lucy deconvolution.

\subsection{{\sl Chandra}}\label{ss:chandra}

Our approach for searching for X-ray emission from the inner knot with {\sl Chandra} involved a special use of both the high time resolution and high spatial resolution of the High-Resolution Camera for Spectroscopy (HRC-S) array in timing mode. 
So doing combines the best spatial resolution and the best time resolution afforded by the {\sl Chandra} instrumentation. 
To achieve the time resolution, it is vital that the counting rate be below the rate where telemetry is saturated \citep[see, e.g.,][]{Tenn01}.
Consequently, the Low-Energy Transmission Grating (LETG) was inserted into the optical path to reduce the flux. 
As inserting the grating is of itself insufficient to avoid telemetry saturation, the trigger threshold of the HRC-S was was raised to reduce the rate of detected events.
This approach was first successfully used in a trial experiment (ObsID 11245) on 2010 November 16.

The next set of observations in this mode was initiated by us on 2013 March 5 as part of a sequence triggered by a $\gamma$-ray flare \citep{Ojha13}. 
Unfortunately, the gain of the HRC had dropped since 2010. 
Consequently, the trigger threshold was too high during our first observation on March 5-6, reducing the counting rate from the pulsar to about a quarter of that expected, dramatically decreasing the number of detected counts.
The next HRC observation was to take place on March 10 and the observation had already been fully planned with the commands to execute onboard the spacecraft. 
To replan at this stage would require an extremely short turnaround, yet it was accomplished. 
Finally, beginning on 2013 October 22, we triggered another sequence of HRC observations in response to the announcement of another $\gamma$-ray flare \citep{Buso13}. 

We also consider ObsID 9765 taken on 2008 January 22. 
For this observation the HRC was operated in a more standard configuration with the threshold set at the nominal value. 
However, to reduce the count rate, a blade was inserted into the optical path, which is no longer allowed. 
This blade also reduced the diffracted flux in one direction. 
By chance, this reduction is greatest near the nominal location of the knot. 
This reduction, together with a long integration time for this observation, results in ObsID 9765 providing the best data for study of the knot.
Table~\ref{t:chandraobs} summarizes the relevant {\sl Chandra} observations.

\section{Properties of the Knot} \label{s:prop}

\subsection{Keck}\label{ss:keckprop}

\begin{figure}[ht]
\begin{center}
\includegraphics[angle=0,width=1\columnwidth]{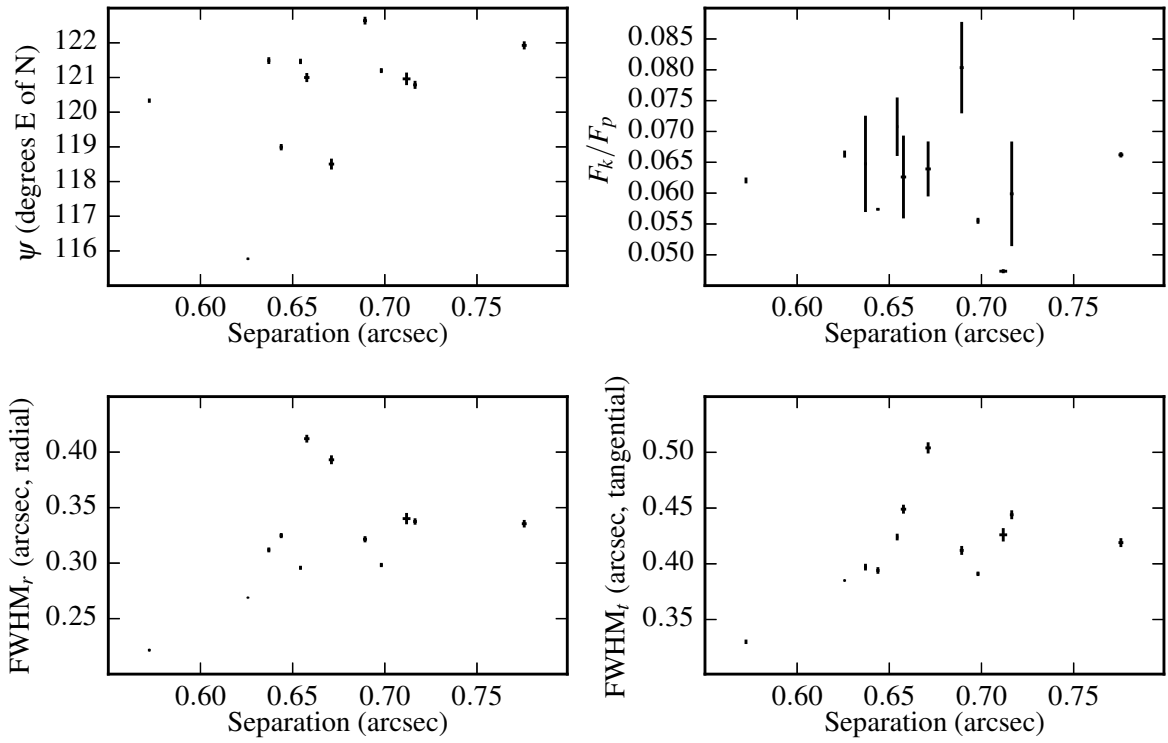}
\figcaption{Variation of the Keck-measured knot properties versus separation of the inner knot from the pulsar.
\label{f:keckknotvar}}
\end{center}
\end{figure}

Table~\ref{t:keckprop} lists the best-fit values of the model parameters for the inner knot for each of the 12 Keck observations based upon the analysis and modeling described in the Appendix. 
The properties are these: $\psi_0$, the position angle (E of N) from the pulsar to the center of the knot; $r_0$, the projected separation between the pulsar and the knot; FWHM$_r$, the full width at half maximum of the knot in the radial direction; FWHM$_t$, the FWHM of the knot in the tangential direction; and $F_k/F_p$, the ratio of the flux in the knot to that in the pulsar.
The flux ratio is used as measuring the fluxes of the 5 stars in in the NIRC2-wide camera's field of view shows that those fluxes change relative to each other from one observation to the next.
This occurs because the nebular background varies spatially across the detector in a non-trivial way,
making it difficult to determine precisely the flux of any given source.
Consequently, we measure the ratio of the knot's flux to the pulsar's, rather than the knot's absolute flux.
Figure~\ref{f:keckknotvar} plots these properties versus the pulsar--knot separation, based upon the Keck data.

We seek possible correlations of these properties with the separation $r_0$ from the pulsar, of the following form:
$p(r_0) = p(r_1) [r_0/ r_1]^q$.
We choose a reference distance $r_1 = 0.6692\arcsec$, obtained from the logarithmic average of the 12 values of $r_0$. 
The position angle of the knot, $\psi_0$, was fit to the linear regression 
$\psi_0 = \psi_0(r_1) +\psi_0^\prime(r_1) [r_0-r_1]$.
Table~\ref{t:keckregression} tabulates the results of the regression analyses of the inner knot’s intrinsic properties with its (projected) separation from the pulsar. 
Based upon the F-test probability, the power-law regression analysis finds marginally significant correlations of the radial width, tangential width, and peak intensity (surface brightness) with separation.
The correlation of the knot's flux to the separation is not significant.

\subsection{HST}\label{ss:hstprop}

Table~\ref{t:hstsvdprop} lists the best-fit values of the model parameters for the inner knot for each of the 17 HST observations based upon the modeling and analysis described in the Appendix.
The properties are these: the position angle $\psi_0$ of the knot with respect to the pulsar, the (projected) distance $r_0$ from the pulsar, radial width (FWHM$_r$) and tangential width (FWHM$_t$) of the knot, the peak surface brightness $S_k$, the knot flux $F_k$, the pulsar flux, $F_p$, and flux ratio $F_k/F_p$.
Although we expected the pulsar flux to be steady, we measured a small ($2.5\sigma$) variation.
Consequently, we prefer to work with the flux ratio $F_k/F_p$. 

\begin{figure}[ht]
\begin{center}
\includegraphics[angle=0,width=1\columnwidth]{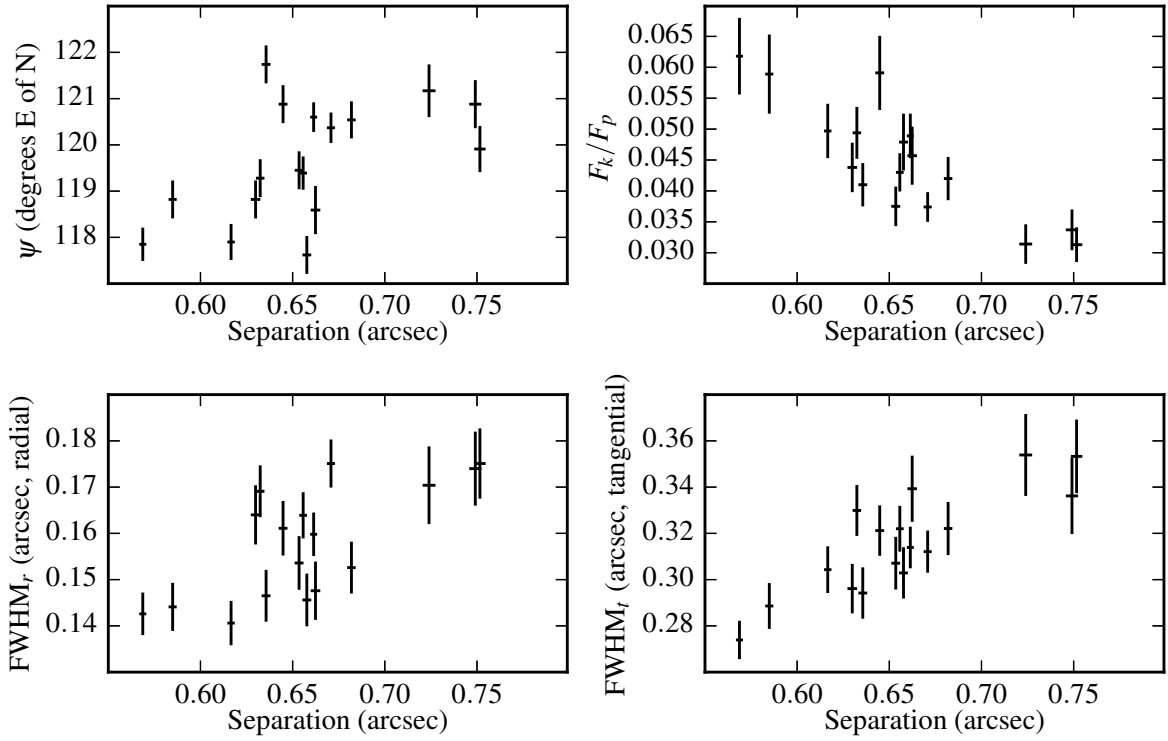}
\figcaption{Variation of the HST-SVD-measured properties with separation of the inner knot from the pulsar.
\label{f:hstknotvar}}
\end{center}
\end{figure}

In the Appendix we discuss two ways of analyzing these data. 
Table~\ref{t:hstsvdprop} presents the results for one of these methods.
However, our general conclusions concerning the knot's optical properties are for the most part the same for either analysis.
Figure~\ref{f:hstknotvar} plots the knot properties versus the pulsar--knot separation, based upon the HST data.

As with the Keck data, we perform a regression analysis of the inner knot’s intrinsic properties with its (projected) separation from the pulsar (Table~\ref{t:hstregression}). 
We choose a reference distance $r_1 = 0.6559\arcsec$, obtained from the logarithmic average of the 17 measured values of $r_0$. 
The position angle of the knot, $\psi_0$, was fit to a linear regression 
$\psi_0 = \psi_0(r_1) +\psi_0^\prime(r_1) [r_0-r_1]$.

Based upon the F-test probability, the power-law regression analysis finds statistically significant correlations of the tangential and radial widths with separation from the pulsar, as well as the previously known \citep{Moran2013MNRAS.433.2564M} anti-correlations of flux and peak surface brightness with separation.
The correlation of $\psi_0$ with separation $r_0$ is the least strong. 
There is a small systematic bias in the power-law index of the FWHM$_r$ correlation, in that the radial width of the knot at times approaches the resolution limit of the instrument and analysis procedures. 
Applying our analysis procedures (see Appendix) to simulated knots of known true radial width, we estimate that the smallest measured values of FWHM$_r$ are oversized by about 7\% and the largest by about 3\%, which would increase the power-law index for the FWHM$_r$--$r_0$ correlation from $q = 0.72$ to about $0.8$. 
Finally, we emphasize that we find similarly strong correlations---except for the flux---using the more traditional HST analysis methods described in the Appendix.

\subsection{{\sl Chandra}}\label{ss:chandraprop}

For the {\sl Chandra} data, the main question is whether X-ray emission from the knot can be detected---especially in the vicinity of the bright, X-ray-emitting pulsar.
The answer, unfortunately, is ``No'' and here we set an informative upper limit to the X-ray flux of the inner knot relative to that of the pulsar. 

Table~\ref{t:chandraprop} shows the number of counts in each of three extraction regions: (1) centered on the approximate location of the optical knot, (2) centered on the pulsar, and (3) placed on a reference region to estimate the background. 
Details may be found in the Appendix and are illustrated in Figure~\ref{f:chandra1}. 
The analysis was performed both for the full phase-averaged data and for the data at pulse minimum.
Consider, for example, the phase-averaged data for ObsID 11245.
Comparing the inner knot counts to the background (Table~\ref{t:chandraprop}) we find an excess of 167 counts. 
If the excess were really due to the knot, we would accordingly expect an excess of 50 ($=167\times0.3$) counts at phase minimum. 
As this was not the case, we attribute the excess to the wings of the PSF from the pulsar and posit that it is due to the spring/fall asymmetry discussed in the Appendix. 
As the impact of that asymmetry is below the statistical noise for the data taken at phase minimum, we then use the phase-minimum data to derive upper limits. 
We convert our phase-minimum upper limit into a phase-average upper limit by accounting for the $0.3$ phase duration of the minimum and then divide by the phase average pulsar flux.
Results are listed in the last column in Table~\ref{t:chandraprop}.

\section{Comparison amongst observations of the inner knot} \label{s:compkhc}

Our {\sl Chandra} observations were unable to detect X-ray flux from the knot, setting a 3-sigma upper limit to the ratio of knot flux to pulsar flux $F_k/F_p < 0.0022$.
This upper limit is 20--40 times smaller than the ratio measured in the optical ($F_k/F_p = 0.044\pm 0.001$) and in the near-IR ($F_k/F_p = 0.079\pm 0.005$), indicating that the optical--X-ray spectral index of the knot is significantly steeper (by at least 0.5) than that of the pulsar. 

Both our infrared and optical observations detected time variability in the pulsar--knot separation, with roughly consistent separations when measured contemporaneously (see the lower portion of Figure~\ref{f:fvsmjd}).
Correlations of knot properties with pulsar--knot separation are very strong for the optical measurements but rather weak for the infrared measurements. 
We attribute this difference to a combination of factors---including poorer statistics for the much shorter NIR measurements and a number of systematic effects, resulting in part from the NIR PSF and background varying across the field and in time.
As the lower portion of Figure~\ref{f:fvsmjd} and Table~\ref{t:keckprop} indicate, 
NIR-measured separations at essentially the same epoch exhibit a scatter substantially larger than the quoted statistical errors in the measurement, thus suggesting the presence of unaccounted systematic errors.

\section{Comparison with $\gamma$-ray fluxes} \label{s:comp}

It is clear from the previous sections that the intrinsic properties of the optical knot are correlated with its projected separation from the pulsar.
The correlation for the NIR knot is but marginally significant. 
Thus, in searching for any relationship between the knot and the $\gamma$-ray flux, we concentrate on the separation versus $\gamma$-ray flux.
The analysis performed to reduce the {\sl Fermi}/LAT data is the same as described in \cite{Maye13}.

\begin{figure}[ht]
\begin{center}
\includegraphics[angle=0,width=1.0\columnwidth]{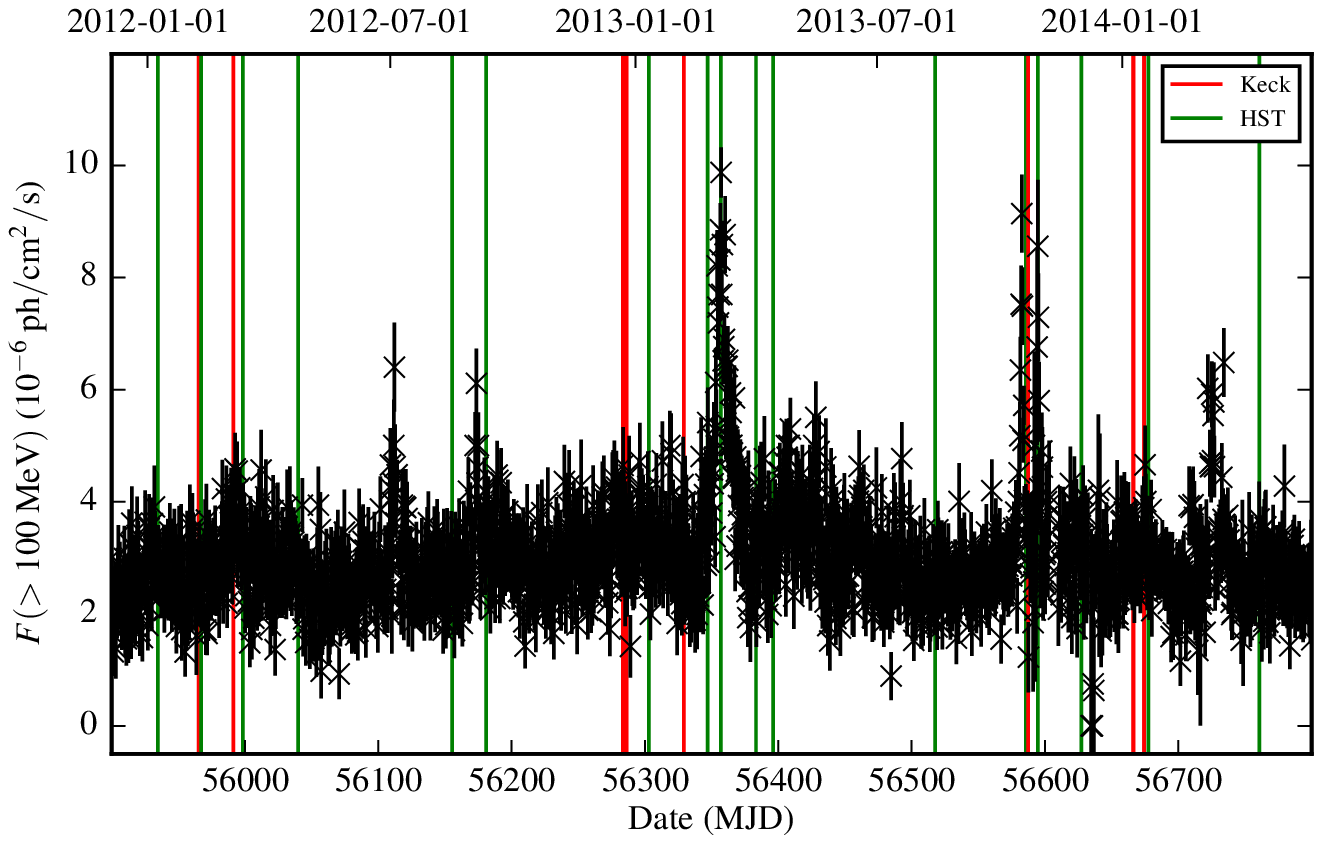}
\includegraphics[angle=0,width=1.0\columnwidth]{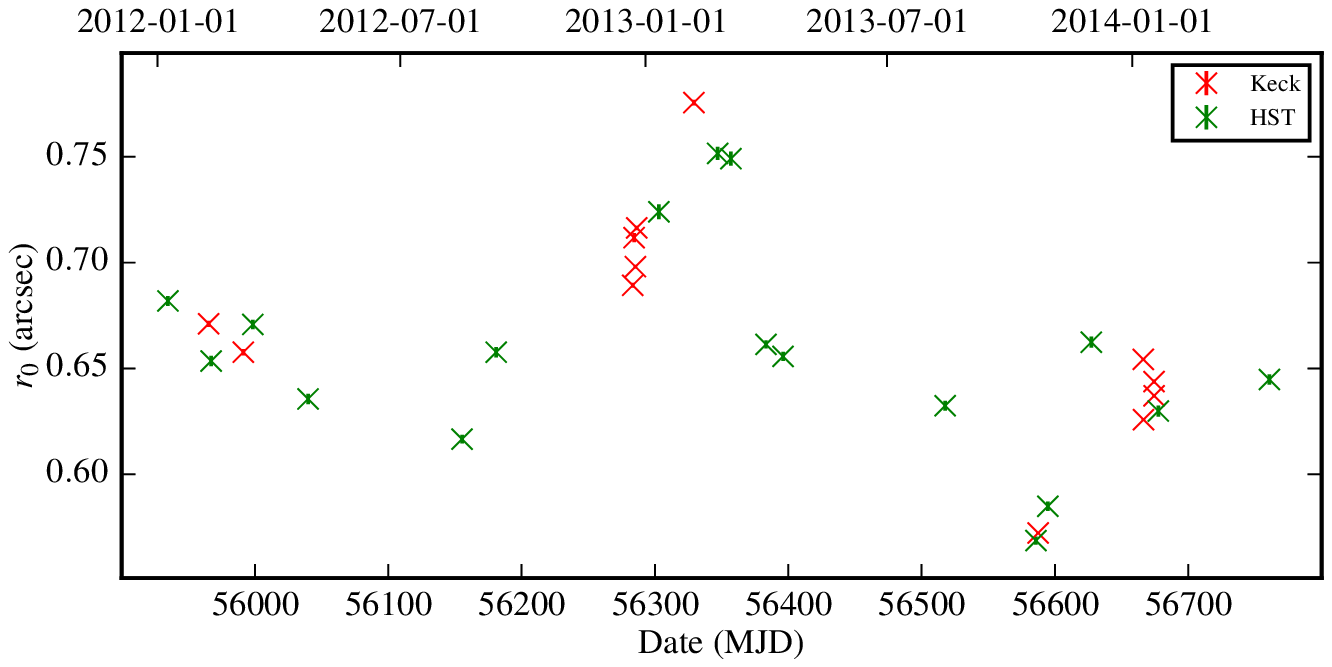}
\figcaption{{\sl Fermi}/LAT 12-hr average flux (upper panel) and pulsar--knot separation (lower panel) versus time, for Keck (red) and HST (green) observations. 
\label{f:fvsmjd}}
\end{center}
\end{figure}

\begin{figure}[ht]
\begin{center}
\includegraphics[angle=0,width=0.9\columnwidth]{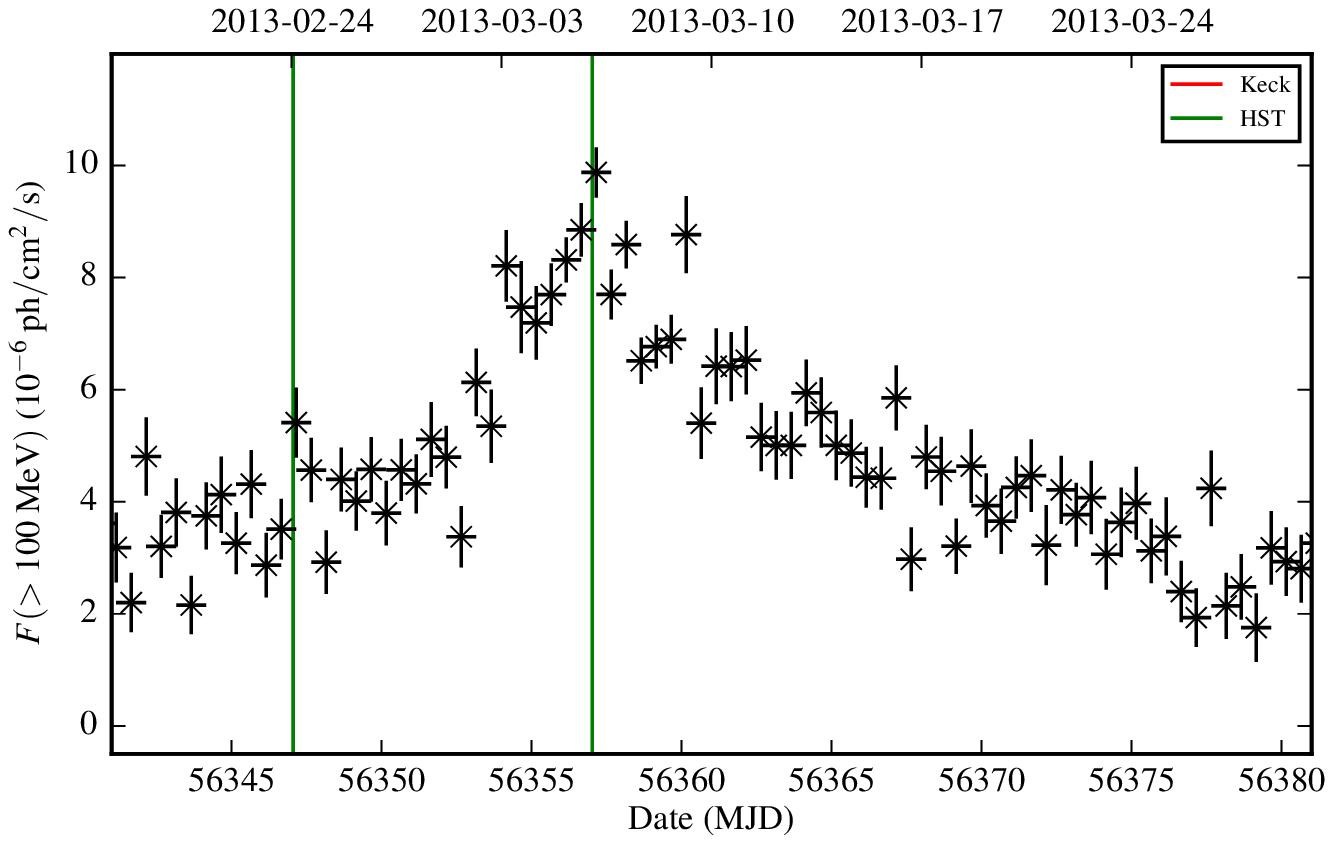}
\includegraphics[angle=0,width=0.9\columnwidth]{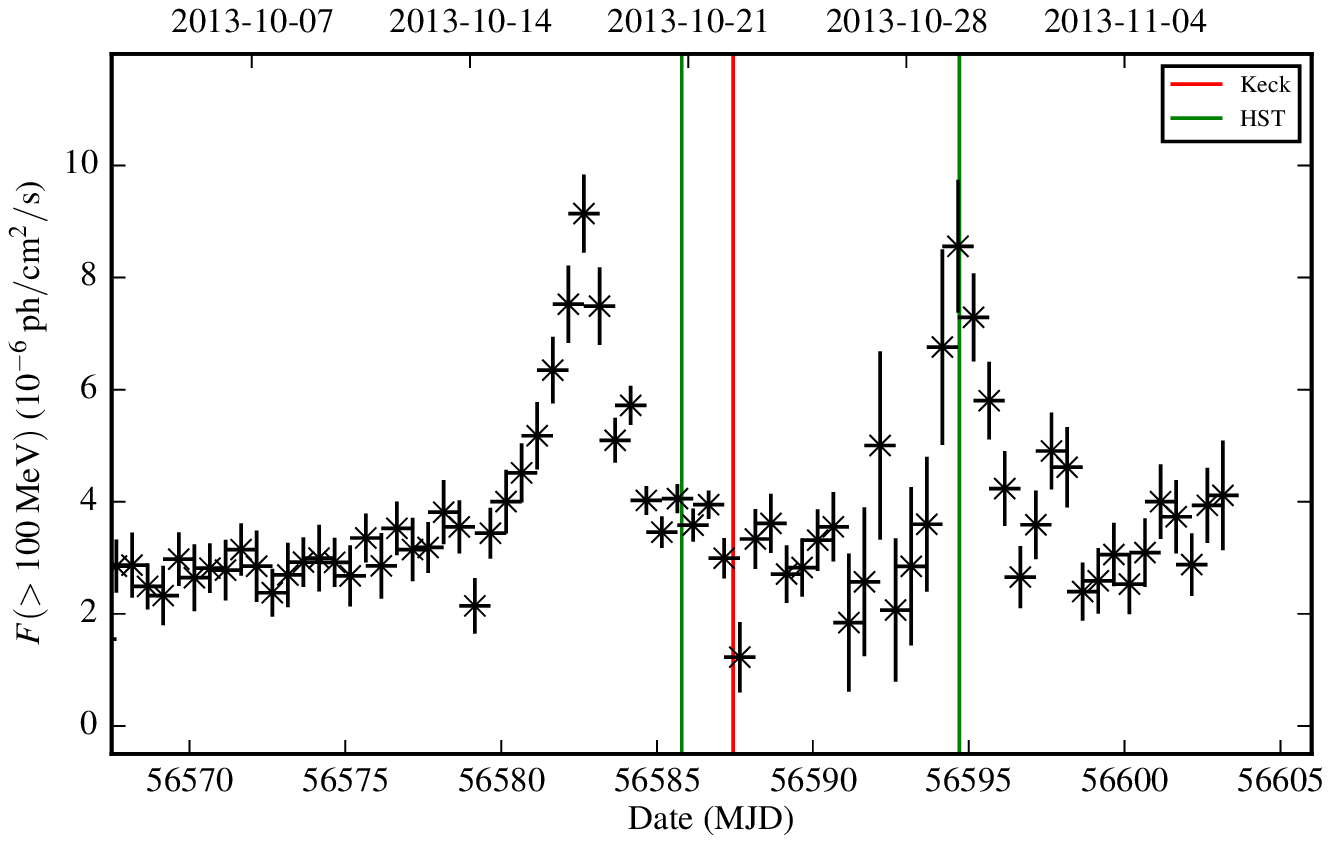}
\figcaption{{\sl Fermi}/LAT 12-hr average fluxes during  Keck (red) and HST (green) observations at the time of the largest $\gamma$-ray flares in 2013 March (upper panel) and in 2013 October (lower panel). 
\label{f:fermi1}}
\end{center}
\end{figure}

Table~\ref{t:fermi} lists the dates and separations measured in the Keck and in the HST observations, followed by the corresponding 12-hour-average {\sl Fermi}/LAT $\geq 100$-MeV fluxes. 
The upper portion of Figure~\ref{f:fvsmjd} shows the {\sl Fermi}/LAT flux as a function of time covering the interval when our Keck and HST observations were made. 
The vertical lines identify times when observations were made with Keck (red) or with HST (green). 
The lower portion of Figure~\ref{f:fvsmjd} plots the pulsar--knot separations on the same time scale.
Figure~\ref{f:fermi1} expands Figure~\ref{f:fvsmjd} at the times of the largest $\gamma$-ray flares.

\begin{figure}[ht]
\begin{center}
\includegraphics[angle=0,width=1\columnwidth]{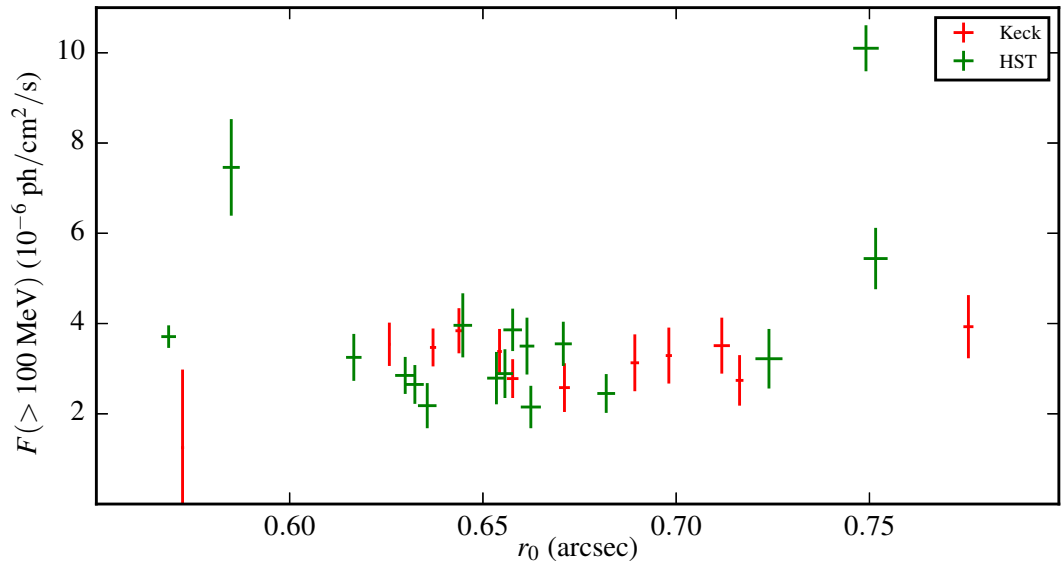}
\figcaption{{\sl Fermi}/LAT 12-hr-average fluxes centered on the times of Keck (red) and HST (green) measurements of pulsar--knot separations.
\label{f:Fvsr0}}
\end{center}
\end{figure}

Figure~\ref{f:Fvsr0} plots the $\gamma$-ray flux versus the measured radial separation of the knot from the pulsar. 
The results captured in this figure clearly do not represent a significant simple correlation of the knot properties with the $\gamma$-ray flux: The formal correlation coefficient is only 0.2. 

Inspection of the $\gamma$-ray light curve and the time series of pulsar--knot separations (Figure~\ref{f:fvsmjd}) indicates that the most energetic $\gamma$-ray flares occurred at a time which is not precisely coincident with the time of greatest pulsar--knot separation. 
It is reasonable to ask how probable such a coincidence would be as a chance occurrence. 
A simple answer is that the probability of a chance coincidence is just the difference in the central times of the two events divided by the total duration of this observing program, as described below.

We next estimate the time and duration of a $\gamma$-ray flare and the excursion in $r_0$ (lower portion of Figure~\ref{f:fvsmjd}), using a simple Gaussian model in both cases and including a linear baseline trend for the $\gamma$-ray flare.
This is not to suggest that a Gaussian properly describes the underlying physical processes, but simply to enable comparison of the two time series, one for the flares, one for the separations.
Fitting separation data in the MJD range 56180-56400 and $\gamma$-ray fluxes in the range 56345-56370 places the midpoint of the inner knot excursion at 56331.8, followed by the $\gamma$-ray flare at 56357.3, 25.5 days later. 
The Gaussian width is a measure of the event duration with FWHMs of 70.7 and 7.6 days for the separation excursion and $\gamma-$ray flare data respectively.

The time between our first and last pulsar--knot separation measurements is 826.1 days.
However, on 4 occasions the gap between separation measurements was longer than the 70.7 day duration of the observed separation excursion. 
We assume then that a separation excursion would be observable if an observation fell within the FWHM window and not observable if it fell outside the window. 
To obtain the total observing program under this assumption, we remove the time of 4 gaps (140.1 days) that exceeded 70.7 days making the total duration of the observing program 686 days.
Thus an estimate of the chance probability is $25.5/686=3.7\%$, which is approximately at the $2\sigma$ level.

It is worth noting that the largest pulsar--knot separation incursion (toward the pulsar) occurs near the second most energetic $\gamma-$ray flare. 
However, the knot separation data are sparser at this time. 
Therefore, we use the previous FWHM value of 70.7 days to establish the time of the incursion. 
With this prior, fitting the separation data in the MJD range 56390-56670 and $\gamma$-ray fluxes in the range 56575-56590 places the midpoint of the inner knot excursion at 56566.0, followed by the $\gamma$-ray flare at 56582.6, 16.6 days later. 

In this second case, the comparable chance probability estimate is $16.6/686=2.4\%$ ($\sim2.5\sigma$). 
As each variation of the knot-pulsar separation is of the opposite sign, we must multiply the probabilities by a factor of 2 when combining them. 
So our estimate of the chance probability that the 2 largest knot excursions/incursions would occur within the observed intervals of the 2 largest $\gamma$-ray flares is $0.037\times0.024\times4=0.4\%$, which is approximately at the $2.9\sigma$ level. 
There are additional $\gamma$-ray flares; however, the third most energetic flare occurs only 12 days later than the one on MJD 56582.6 and therefore is well within the assumed 70-day range of the associated knot incursion.  
The other flares are significantly weaker and less well-sampled by the optical/infrared observations.

Although coverage is sparse, there is no compelling evidence in the pulsar--knot separations for large ($>0.01\arcsec$) excursions from the mean separation at time-scales comparable to the flare durations. 
Indeed, knot incursion/excursion time scales appear to be $>10\times$ longer than flare durations, so direct correlation of the $\gamma$-ray light curve with the separation time-evolution produces correlation coefficients that are rather difficult to interpret without associated simulations. 
Such simulations would require assumptions regarding an underlying physical model relating the two curves. 
As we have no such model, any correlation analysis would be purely {\it ad hoc}.

The 0.4\% value for the chance probability is based upon a rather simplistic analysis involving only 2 flares and was, of course, determined {\it a posteriori}. 
However, in continuing to monitor the Crab at multiple wavelengths with the goal of understanding the origin of its $\gamma$-ray flares, the inner knot will continue to be a feature of interest.  
Based upon the $r_0$ time scale of 71 days (30-day Gaussian $1\sigma$), a monitoring program which includes monthly optical observations is necessary to detect significant inner-knot excursions from its mean pulsar separation. 
Monitoring every $\sim3$ weeks should be sufficient to characterize fully these variations and to verify the small statistical sample from which these parameter estimates were derived.

We also note that an inner-knot incursion with an $\approx2$-month FWHM timescale and $0.075\arcsec\pm 0.025\arcsec$ pulsar--knot separation inward amplitude was also reported by \citet{Moran2013MNRAS.433.2564M}. 
This event appears to be quite similar in duration and amplitude to our observations. 
The reference also reports an $\approx2\sigma$ apparent change in the knot polarization parameters, which occurred $22\pm 5$ days after the time of maximum incursion. 
The degree of linear polarization steps from $59.9\% \pm 6.9\%$ on 2005 Nov 25 to $42.8\% \pm 6.2\%$ 2005 Dec 5 and the position angle changes from $125.4\degree\pm 3.3\degree$ to $121.6\degree \pm 4.2\degree$. 
Interestingly, the time frame of this polarization shift occurred in a window closely matching both the 25.5 and 16.6 day $\gamma$-ray flare delays after the peak inner-knot excursion/incursion times discussed above. 
While the claimed $2\sigma$ result is not very significant, we observe that prior to the step near the start of 2005 December, all 10 of the reported measurements of the degree of linear polarization fall into a population with mean $60.7\%$ and small standard deviation $1.0\%$, whereas the 2 subsequent measurements had a mean of $43.4\%$ with standard deviation of $0.8\%$.  
If we suppose that the quoted errors in the reference apply to the absolute, but not relative measured values, then the size of the step change would be $>10\sigma$ instead of $\approx 2 \sigma$. 
This may be a reasonable assumption, as the point-to-point measurement variations are smaller by a factor $>6$ than the quoted errors. 
Similar arguments would apply to the polarization position angle. 
From this we conclude that polarization should also be of interest in future attempts to relate inner-knot behavior to $\gamma$-ray flaring.

\section{Implications for Theoretical Models}  \label{s:implications}

\begin{figure}[hbt]
\begin{center}
\includegraphics[width=\textwidth]{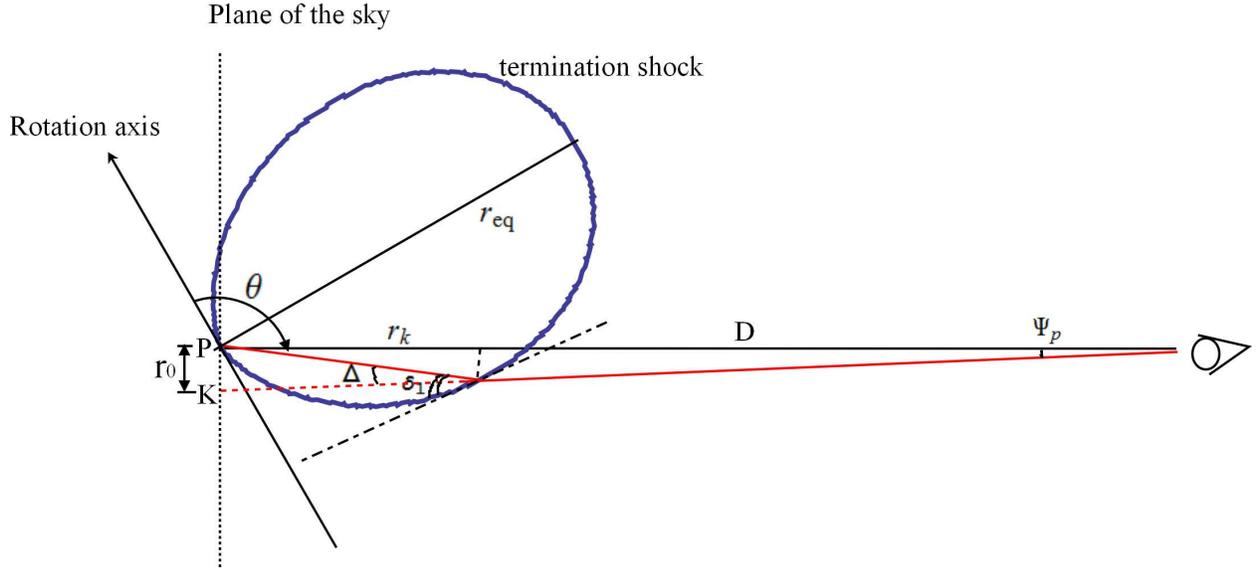}\\
\caption{Geometry of the termination shock on the meridional plane determined by the pulsar rotation axis and the observer's line of sight toward the pulsar. 
The pulsar is located at point P. 
The angle between line of sight and pulsar rotation axis is $\theta_{\rm ob}=2\pi/3$, consistent with that measured by \citet{Ng08}. 
The knot is projected on the plane of the sky at an angle $r_0\approx0.65''$ southeast of the pulsar (represented by point K here). 
The angle between the upstream velocity and the shock surface is $\delta_1$. 
The outflow is deflected from the radial direction by an angle $\Delta$. 
\label{fig:shock}}
\end{center}
\end{figure}

In the basic model of the Crab Nebula \citep[e.g.,][] {ReesGunn1974MNRAS.167....1R}, the pulsar wind passes through a shock at a radius $\sim3\times10^{17}$~cm where the wind momentum flux balances the nebular pressure.
However, the wind should be anisotropic.
For example, the momentum flux in a split-monopole $\propto\sin^2\theta$ \citep{Bogovalov1999A&A...349.1017B} and recent simulations suggest $\propto\sin^4\theta$ \citep{Tchekhovskoy2013MNRAS.435L...1T, Tchek15} where $\theta$ is the polar angle.
As a result, the shock is likely to be quite oblate.
The sections of the shock near the poles would be oblique and much closer to the pulsar than the equatorial part.
The observed radiation presumably comes from relativistic particles (electrons or positrons) accelerated behind the shock.
Since the outflow from an oblique shock can remain relativistic, we would be able to see a compact, emitting feature (the inner knot) if the relativistic outflow happens to be aligned with our line of sight so that its emission is beamed.
The inner knot should appear to us as having some offset from the pulsar due to the deflection of the outflow from its presumed radial motion (Figure \ref{fig:shock}).
If the shock is approximately axisymmetric as one would expect, the projected emitting site would fall on the projected symmetry axis, leading to an alignment with the jet.
The shape of the inner knot should also be more or less symmetric about the axis; whether it is elongated parallel or perpendicular to the axis depends on the geometry of the oblique shock.
Furthermore the synchrotron emission should be linearly polarized \citep[] [Yuan \& Blandford 2015 in preparation] {KomissarovLyutikov2011MNRAS.414.2017K}.

The shock model predicts that the scaling between the observed properties of the knot---projected pulsar--knot separation $r_0$, tangential angular width FWHM$_t$, radial angular width FWHM$_r$ and surface brightness $S_k$---should be determined by the upstream magnetization $\sigma\equiv B_1^2/\mu_0n_1\gamma_1^2mc^2$ (where $m$ is electron mass and $B_1$, $n_1$, $\gamma_1$ are upstream magnetic field, proper density, bulk Lorentz factor, respectively) plus three more parameters that characterize the shape of the shock near the emitting site: 
The unprojected pulsar-knot separation $r_k$; the incident angle $\delta_1$, defined to be the angle between upstream velocity and the shock surface as shown in Figure \ref{fig:shock}; and the shock poloidal radius of curvature $R_c$.

Let the outflow be deflected from the radial direction by an angle $\Delta$ at the shock. 
We then have $r_0=r_k\Delta$. 
The knot tangential size is determined by the Doppler beaming and the shock radius of curvature in the toroidal direction as FWHM$_t\approx 2 r_k/\gamma$, where $\gamma$ is the downstream Lorentz factor. 
In the simplest shock model with an isotropic plasma, both $\Delta$ and $\gamma$ are functions of $\sigma$ and $\delta_1$:
$\Delta=\delta_1-\arctan(\chi\tan\delta_1)$ and $\gamma=1/(\sin\delta_1\sqrt{1-\chi^2})$, where $\chi\equiv v_{2\perp}/v_{1\perp}=B_1/B_2=(1+2\sigma +\sqrt{16\sigma ^2+16\sigma +1})/6(1+\sigma )$ is the compression ratio at the shock, $v_{1,2\perp}$ being the component of upstream/downstream velocity perpendicular to the shock \citep{KomissarovLyutikov2011MNRAS.414.2017K}.
The radial knot size FWHM$_r$ should also be proportional to $r_k/\gamma$ but will have an additional factor involving the ratio $R_c/r_k$.

As to the intensity, we assume that the emitting particles have a power-law distribution $\propto n'\gamma'^{-p}$ in the fluid rest frame, where $n'$, $\gamma'$ are particle density and Lorentz factor in the same reference frame, and $p\sim2.6$ from the measured spectral index of the knot \citep{Soll03, Melatos05}.

The emissivity in the fluid rest frame is $j'_{\nu'}(\nu')\propto n'
{B'}^{(p+1)/2}{\nu'}^{-(p-1)/2}$, where $B'$ and $\nu'$ are magnetic field and radiation frequency measured in this reference frame.
After transforming to the nebula frame, the emissivity has a dependence on the flow Lorentz factor $\gamma$, particle (improper) number density $n$, and magnetic field $B$ as follows: 
$j_{\nu}\propto \mathcal{D}^{2+(p-1)/2}(n/\gamma) (B/\gamma)^{(p+1)/2}$, 
where $\mathcal{D}=\nu/\nu'=\sqrt{1-\beta^2}/(1-\vec{\beta}\cdot\vec{n})$ is the Doppler factor, in which $\vec{\beta}$ is the flow velocity and $\vec{n}$ is a unit vector along the direction of line of sight.
The surface brightness corresponds to the emissivity integrated along the line of sight in the nebular frame: $S_{\nu}=\int j_{\nu} dl$.
To estimate the peak surface brightness, one takes the maximum Doppler factor $\mathcal{D}\approx2\gamma$, and we get $S_{\nu,peak}\propto n B^{(p+1)/2}\ell$ where $\ell$ is an estimation of the length of the emitting region along the line of sight.
As roughly $B\propto r_k^{-1}$ and $n\propto r_k^{-2}$, we have $S_{\nu,peak}\propto r_k^{-2-(p+1)/2}\ell$.

These previous relations allow us to set the constraints discussed below using both the steady and variable properties of the knot.

\subsection{Variability}

In the shock model, variability can arise either upstream or downstream of the shock.
Firstly, stress tensor variation in the nebula can cause the shock radius and shape to vary over time.
Three dimensional MHD simulations \citep{Porth14} show that even if the upstream condition is fixed, the post-shock flow can be quite variable: 
there is vortex-shedding from the termination shock, and the shock constantly interacts with waves and vortices in the nebula.
Those variations are expected to be mostly magnetic in origin, and the interactions between the shock and the downstream flow are quite nonlinear.
The picture is similar to earlier 2D simulations \citep[e.g.,][] {Camus09} except that the kink instability sets in so that the hoop compression is less prominent as in the 2D case; also short-term variability is less pronounced.
The 3D simulations give a typical variation time scale of less than one year.

Now suppose that the change of downstream pressure causes the shock radius $r_k$ to change. 
For simplicity we first assume that $\Delta$, $\ell$, $\gamma$ and $p$ stay more or less the same, then from the above scaling relations, we have $S_{\nu,peak}\propto r_0^{-2-(p+1)/2}$, FWHM$_t \propto r_0$, and knot flux $F_{\nu} \propto S_{\nu,peak}{\rm FWHM}_t\, {\rm FWHM}_r \propto r_0^{-(p+1)/2}$, assuming FWHM$_r$ to be also roughly proportional to $r_k$. 
In reality $\Delta$, $\ell$, $\gamma$ change with the shock radius $r_k$ as well, as the shock shape and upstream incident angle $\delta_1$ vary. 
Consequently, the observed correlation is more complicated and less clean; however, the general trend is consistent with our expectation of the shock model.

Variations initiated by the pulsar can also cause the knot properties to change.
In such a scenario, the shortest possible variation time scale is $t_v\sim r_k(1-\cos (1/\gamma))/c\sim r_k/(2\gamma^2c)\sim 1\ \text{day} (r_k/10^{17}\, \text{cm})(\gamma/5)^{-2}$ up to 10\% inaccuracy assuming $\gamma\gtrsim3$.
However, there is no evidence in the pulsar timing that this is actually happening.

\subsection{Time averaged properties of the knot}
The properties of the knot may be used to set constraints on the flow composition and how much dissipation occurs at the shock.
One constraint comes from geometrical relations amongst the three measured angles $r_0$, FWHM$_t$, and FWHM$_r$.
For the simplest shock model, we have $r_0=r_k\Delta$ and FWHM$_t\approx r_k/\gamma$, where $\Delta$ and $\gamma$ only depend on $\sigma$ and $\delta_1$, such that the ratio FWHM$_t/r_0$ sets limit on $\sigma$ and $\delta_1$. 
We find that FWHM$_t/r_0=2/(\gamma\Delta)=2\sin\delta_1\sqrt{1-\chi^2}/[\delta_1-\arctan(\chi\tan\delta_1)]\ge2.8$, with the minimum at $\sigma=0$ and $\delta_1=0.39$. 
In contrast, the observation gives FWHM$_t\sim0.5 r_0$, indicating that some important ingredient is missing in the standard shock model.
For example, the upstream flow might have already been bent before arriving at the shock; and/or the anisotropic particle distribution downstream might be important in determining the size of the knot.
Although more detailed investigations will be described elsewhere, the above discussion on variability and the following discussion on energetics should be valid for quite general shock models.

The shape of the knot also has some interesting implications.
In both the {\sl Hubble} and Keck images, the knot is not just a simple ellipse that only involves quadratic terms of coordinates; it has curvature and appears to be convex away from the pulsar (``smile''), so at least a third moment of the coordinates is needed to describe the shape.
A simple intuitive picture is based on the geometry that the shock surface is similar to a torus centered on the pulsar and we look through the hole from the bottom. 
If emission from the shock outflow roughly follows the local toroidal magnetic field, one would imagine the knot to be convex toward the pulsar (``frown'')---contradictory to what we see.
In fact, we should not overlook the possibility that the outflow may have significant bending downstream.
If the outflow bends toward the equator by an angle $\sim1/\gamma$ within the emission length, we would see the knot more extended (contracted) on the far (near) side with respect to the pulsar, as additional flow lines enter (originally aligned flow lines leave) the emission cone on the corresponding side.

Another constraint comes from the observed flux of the knot.
In the $K'$ band, the typical knot-pulsar flux ratio is $\sim0.063$ (Table~\ref{t:keckprop}).
If we take the photometry of the pulsar from \citet{SandbergSollerman2009A&A...504..525S}, the pulsar $K_s$ band magnitude is $13.80\pm0.01$, after dereddening with the best fit value $E(B-V)=0.52$
\citep{Sollerman2000ApJ...537..861S} and the extinction law from \citet{Fitzpatrick1999PASP..111...63F}, this corresponds to a flux of $2.2\times 10^{-26}\text{erg}\, \text{s}^{-1}\text{cm}^{-2}\text{Hz}^{-1}$.
So the inner knot flux in $K'$ band should be roughly $F_{\nu}^{\text{knot}}=1.4\times 10^{-27}\text{erg}\, \text{s}^{-1}\text{cm}^{-2}\text{Hz}^{-1}$.
Similarly, optical data (Table \ref{t:hstsvdprop}) on average give a dereddened flux of $F_{\nu}^{\text{knot}}=1\times 10^{-27}\text{erg}\, \text{s}^{-1}\text{cm}^{-2}\text{Hz}^{-1}$.
However, we do not attempt here to get a spectral index using the time averaged fluxes due to the variability of the knot and the fact that most of the multi-wavelength observations are not simultaneous.
Previous measurements by \citet{Soll03} and \citet{Melatos05} indicate that the spectrum of the knot in IR to optical range is a power law $F_{\nu}\propto \nu^{-\alpha}$ with $\alpha\sim0.8$, and \citet{Melatos05} further measured that this power law extends to NUV ($160-320$ nm), with a flux in NUV
$F_{\nu}^{\text{knot}}=0.12\,{\rm mJy}=1.2\times 10^{-27}\text{erg}\, \text{s}^{-1}\text{cm}^{-2}\text{Hz}^{-1}$---this is the highest energy at which the knot has been measured.
A few more recent measurements show slightly different spectral index ranging from 0.63 \citep{Tzia09} to 1.3 \citep{SandbergSollerman2009A&A...504..525S} but unfortunately these are based on non-simultaneous data. 
It would be very interesting to refine the spectral measurement using near simultaneous IR/optical images in the future.
In what follows we adopt a fiducial spectral model $F_{\nu}^{\text{knot}}\sim 10^{-27}(\nu/10^{15}\,\rm{Hz})^{-0.8}\text{erg}\, \text{s}^{-1}\text{cm}^{-2}\text{Hz}^{-1}$ with the caveat that this could
change with better observations.

Using the above spectral model, we find that the knot integrated flux in IR/optical band ($(0.1-2)\times10^{15}\,\text{Hz}$) is $F_{\rm O/IR}^{\text{knot}}=2.6\times10^{-12}\,\text{erg}\,
\text{s}^{-1}\text{cm}^{-2}$.
Thus, the luminosity of the knot per steradian is $L_{\text{O/IR}, \Omega}=D^2F_{\text{O/IR} }^{\text{knot}}=10^{32}\,\text{erg}\, \text{s}^{-1}$, where $D\sim2$ kpc is the distance of the Crab.
If we define $\epsilon\equiv L_{\text{O/IR},\Omega}/(\dot{E}/4\pi)$, where $\dot{E}=5\times10^{38}\ {\rm erg}\,{\rm s}^{-1}$ is the pulsar spin-down power, $\epsilon$ can be used as a measure of the radiative efficiency of the shock along the direction of line of sight in IR to optical range. Here we find
$\epsilon=2.5\times10^{-6}$ and the shock is highly adiabatic.

We can get an estimation of the enthalpy fraction of IR/optical emitting particles in the downstream flow \begin{equation} 
\eta(\gamma')=\frac{8}{3}\, \frac{\gamma t'_{\text{cool}}}{t_{\text{flow}}}\,
\frac{\nu  L_{\nu ,\Omega }}{\dot{E}/(4\pi)}=\frac{8}{3}\,\frac{\gamma t'_{\text{cool}}}{t_{\text{flow}}}\,\frac{\nu  L_{\nu ,\Omega }} {\dot{E}/(4\pi)} ,
\end{equation}
where $t_{\text{flow}}$ is the flow time scale and $t'_{\text{cool}}=6\pi \epsilon _0m^3c^3/(e^4B'^2\gamma ')\approx10^2\gamma^2B_{-3}^{-3/2}\nu_{14}^{-1/2}\ \rm{years}$ is the
synchrotron cooling time of the particles in the fluid rest frame (we have adopted $B=B_{-3}\,\rm{mG}$ and $\nu=\nu_{14}10^{14}\,\rm{Hz}$ for the numerical value).
Thus the particle injection rate per unit steradian in corresponding energy band is
\begin{equation}
\dot{N}(\gamma')=\frac{3\eta(\gamma') }{4 \gamma ' m c^2}\frac{\dot{E}} {4\pi\gamma} .
\end{equation}
Adopting typical values $\gamma=5\gamma_5$, $t_{\rm flow}=r_{17}10^{17}\,{\rm cm}/c$, we find that for IR emitting particles, $\eta_{\rm{IR}}=0.05 \gamma_5^3B_{-3}^{-3/2} r_{17}^{-1}$, and $\dot{N}_{\rm{IR}}=2\times10^{36}\gamma_5^2B_{-3}^{-1}r_{17}^{-1}\; \text{s}^{-1}\text{sr}^{-1}$.
The spectral index of the knot indicates that the particle distribution powerlaw index
should be $p\sim2.6$, thus the IR emitting particles should comprise the majority of particle pressure downstream.
From the above estimations, it seems that the particle injection rate at the shock is marginally consistent with the scenario that most of the IR/optical emitting particles are provided by the shock.

Regarding particle acceleration mechanisms, we notice that IR emitting particles usually go through $N=t_{\rm flow}\nu_g$ Larmor orbits, where $\nu_g = \nu_B / \gamma^\prime$ is the relativistic gyrofrequency---the cyclotron frequency ($\nu_B$) divided by the electron's Lorentz factor ($\gamma^\prime$). 
Eliminating $\gamma^\prime$ in favor of the magnetic field ($B$) and observing frequency ($\nu$), the number of orbits within the flow timescale is $N\approx 6\times10^4B_{-3}^{3/2}r_{17}\nu_{14}^{-1/2}$.
To allow sufficient time for acceleration mechanisms requiring instabilities or stochastic processes to operate, requires that $N>>1$. 
However, for particles emitting $\gamma$ rays of energy 300~MeV, $N\lesssim 1$, thus requiring special acceleration mechanisms---if the emitting electrons are indeed accelerated at the shock.

One more constraint comes from the polarization of the knot.
Most recent polarimetry performed by \citet{Moran2013MNRAS.433.2564M} gives a high polarization degree $\sim60\%$, with position angle aligned with the symmetry axis, indicating strongly a toroidal magnetic field.
Here we need to be careful about the depolarization effect due to relativistic kinematics \citep{Lyutikov2003ApJ...597..998L}.
What happens then is that Lorentz transformation of the emission of a relativistic plasma from the comoving frame to the lab frame produces a rotation of the polarization vector.
For a curved emitting surface, neighboring fluid elements have slightly different velocities, thus their polarization vectors experience different amounts of rotation.
As one sums the contribution from the visible surface (this is essentially what we observe), the result is some degree of depolarization.
We find that for the oblique shock in the Crab Nebula, under an ultrarelativistic approximation, the upper limit of polarization degree is similar to that in \citet{Lyutikov2003ApJ...597..998L}: $56.25\%$ for particle spectral index $p=3$ and $43.4\%$ for $p=2$.
Thus, the observed high degree of polarization and its possible variation is highly constraining on theoretical models. It would be interesting to repeat this observation, especially with adaptive optics.

\section{Conclusions} \label{s:conclusions}

We have (1) introduced a new approach to analyzing time series of images; 
(2) discovered that key properties of the knot (radial width, tangential width, flux) are correlated with the time-variable separation between the knot and the pulsar; 
(3) shown that, with the available data, it is not possible to determine a strong correlation between the knot separation and the occurrence of $\gamma$-ray flares; 
(4) set an upper limit to the low-energy X-ray flux from the inner knot; 
(5) discussed the implications of our observations to set constraints on particular elements of the standard shock modeling of the relativistic outflow from the pulsar. 
These include inferences as to wind magnetization $\sigma$, shock shape parameters such as incident angle $\delta_1$ and poloidal radius of curvature $R_c$, as well as the IR/optical emitting particle enthalpy fraction. 
We found that while the standard shock model gives good agreements with observations in many aspects, there remain two puzzles: 
(a) the angular size of the knot relative to the pulsar--knot separation is much smaller than expected; 
(b) the variable, yet high degree of polarization is difficult to reconcile with a highly relativistic outflow. 
We also found that the IR/optical flux of the inner knot is marginally consistent with the scenario that the shock accelerates most of the optical emitting particles in the nebula.

{\bf Acknowledgments}
The \textit{{\sl Fermi}} LAT Collaboration acknowledges generous ongoing support from a number of agencies and institutes that have supported both the development and the operation of the LAT as well as scientific data 
analysis. 
These include the National Aeronautics and Space Administration and the Department of Energy in the United States, the Commissariat \`a  l'Energie Atomique and the Centre National de la Recherche Scientifique / Institut National de Physique Nucl\'eaire et de Physique des Particules in France, the Agenzia 
Spaziale Italiana and the Istituto Nazionale di Fisica Nucleare in Italy, the Ministry of Education, Culture, Sports, Science and Technology (MEXT), High Energy Accelerator  Research Organization (KEK) and Japan Aerospace Exploration Agency (JAXA) in Japan, and the K.~A.~Wallenberg Foundation, the Swedish Research Council and the Swedish National Space Board in Sweden.
Additional support for science analysis during operations phase is gratefully acknowledged from the Istituto Nazionale di Astrofisica and the Centre d'Etudes Spatiales in France.
The research leading to these results has also received funding from the European Commission Seventh Framework Programme (FP7/2007-2013) under grant agreement n. 267251.
Several of the authors would also like to acknowledge both funding and solid support from the {\sl Chandra} X-ray Center and the {\sl Hubble Space Telescope} Science Institute under a number of observing proposals: GO3-14054Z, GO3-14057Z, GO4-15058Z, GO4-15059Z, GO-13109, GO-13196, GO-13348

\appendix
\section{Appendix - Data Analysis}\label{s:da}

\subsection{Keck}\label{ss:da:keck}

To study the knot using the high-resolution Keck adaptive-optics (AO) data, we characterize the region around the pulsar using an empirically measured point spread function and an analytical model of the knot geometry.

\subsubsection{PSF Subtraction}

\begin{figure}[ht]
\begin{center}
\includegraphics[angle=0,width=1\columnwidth]{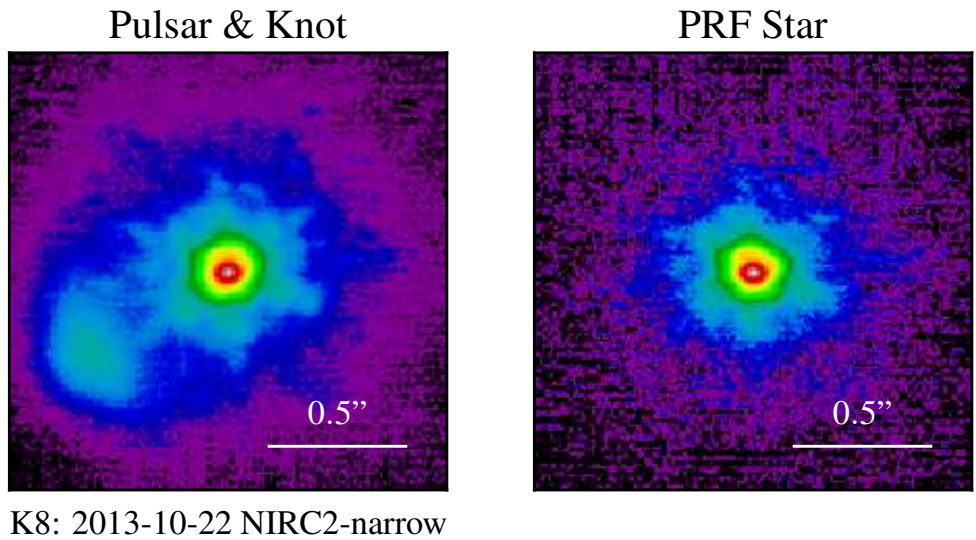}
\figcaption{The Pulsar (left) and a nearby star (right) from Keck data taken on 2013 October 22. (a) An H filter image of the pulsar and knot (in the lower-left) as seen by the NIRC2-narrow camera ($0.01\arcsec$ pixels and a $10\arcsec\times10\arcsec$ field of view). 
The structure of the knot is resolved and separated from the pulsar by a statistically significant valley. 
(b) A nearby ($5\arcsec$) comparison star, used to establish the point-response-function.
Both images are shown with a logarithmic color stretch.
\label{f:keckprf}}
\end{center}
\end{figure}

We subtracted the PSF of the pulsar using a nearby PSF star (Figure~\ref{f:keckprf}). 
During our observations, the laser guide star, and so the center of the anisoplanatic patch, was aimed at the point half-way between the PSF star and the pulsar. 
We fit the PSF star, a background plus a power-law based seeing-disk to the pulsar. 
For the fit, we excluded the region in the pulsar subimage that surrounds the knot, and used the other $270^{\circ}$. 
The background was fit with a sloping plane model, to allow for the variable nebular background in the pulsar and PSF star subimages. 
The fitting was performed with a Levenburg-Marquadt fitter.

\subsubsection{Deconvolution}

We chose not to deconvolve the knot with the empirically measured point spread function. 
Tests on several images showed that deconvolution of the shape of the knot made insignificant differences to the fit values and uncertainty. 

\subsubsection{Knot Model}

To fit the geometric parameters of the knot, we masked out the center diffraction-limited core of the pulsar, and used a Levenburg-Marquadt fitting technique to fit a two dimensional Gaussian to the shape of the Knot.

We used a model of the form

         $S(r,\psi)=S_b + S_k \exp(-\frac{1}{2}((r-r_0)^2/\sigma_r^2 + (\psi-\psi_0)^2/\sigma_\psi^2))$  


We also allowed the center of the coordinate system ($x_0, y_0$) to vary slightly ($0.05\arcsec$). 
To determine the fit uncertainties, we used the $1$-$\sigma$ errors from the self-covariance of the fit parameters. 
Results were tabulated in Table~\ref{t:keckprop}.

\subsection{HST}\label{ss:da:HST}

\subsubsection{HST - Traditional Analysis}\label{sss:da:HST:trad}

In order to measure the knot properties (position, flux, tangential and radial width) we used the SExtractor package \citep{Bert96}.
SExtractor has been extensively used for the analysis of HST data, in particular for the ACS camera \citep[e.g., for the {\sl Hubble} Ultra Deep Field project,][]{Beck06}. 
For source detection, we require a minimum of five contiguous pixels with a detection threshold $5\sigma$ above the root-mean-square (rms) background, with a total of 32 deblending subthresholds, and with a contrast parameter of 0.005, setting the background mesh size to 16$\times$16 ACS pixels. 

The position of a source is evaluated by SExtractor as the barycenter of the source brightness distribution. 
Flux is computed within an elliptical aperture, using an implementation of the method by \citet{Kron80}. 
Note that the pulsar is not saturated and that diffraction spikes from the pulsar that might cross the inner knot only marginally affect its flux measurements. 
The parameters of this ellipse (semiaxes and orientation), evaluated using the second moments of the object's brightness distribution also yield a measure of the object morphology. 
Indeed, for the case of the knot, the direction of the minor axis of the ellipse turned out to be consistent with the pulsar--knot direction in all images. 
Thus, the minor axis and the major axis of the ellipse measure the RMS full widths (FWRMS) in the radial and tangential directions, respectively.

In order to assess systematic errors, we performed simulations with the ESO/MIDAS software\footnote{ https://www.eso.org/sci/software/esomidas/}. 
We added to the ACS images a ``synthetic knot''. 
To generate such an artificial source, we assumed a two-dimensional Gaussian brightness distribution, with the minor axis aligned with the true pulsar--knot direction. 
The synthetic knot was positioned to the NW of the pulsar, opposite but along the true pulsar--knot direction, and at an angular distance comparable to that of the true knot. 
We repeated the exercise by varying the flux, position, and morphology of the synthetic knot and we estimated the uncertainties in the parameters recovered using SExtractor \footnote {http://www.astromatic.net/software/sextractor.}

Table~\ref{t:hsttrad} lists the best-fit values of the model parameters of the inner knot for each of the 17 observations based upon this traditional analysis and modeling.
Table~\ref{t:hsttradregression} shows the results of a regression analysis on the listed parameters.

\subsubsection{HST Singular Value Decomposition (SVD)}\label{sss:da:HST:svd}

As the inner knot is within $1\arcsec$ of the Crab pulsar and varies in position and size, we also developed special procedures for characterizing it using the central $121\times121$-pixel ($6.05\arcsec\times6.05\arcsec$) images of the 17 HST/ACS observations. 
The major steps are (1) to remove the pulsar from each image, (2) to use the residual image to generate an image of the inner knot, and (3) to characterize the properties of the inner knot.

{\bf Remove pulsar and generate an image.}

\begin{figure}[ht]
\begin{center}
\includegraphics[angle=0,width=0.9\columnwidth]{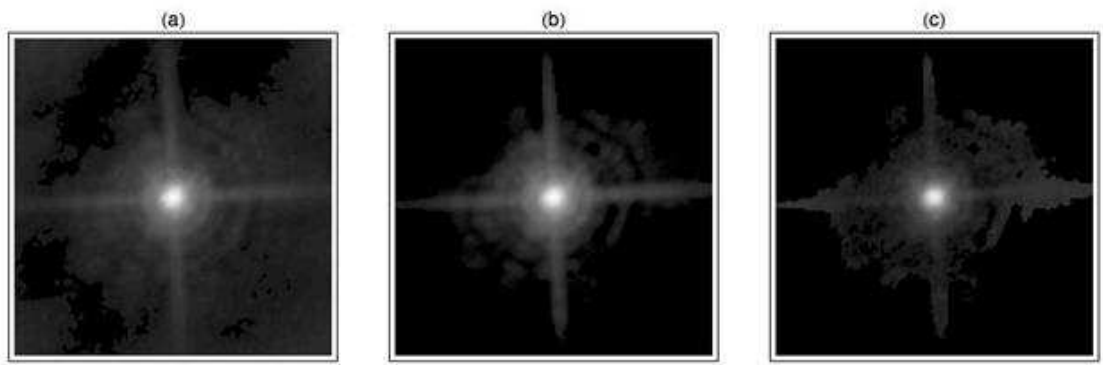}
\figcaption{Images illustrating the steps in determining the HST/ACS PSF for this analysis: 
(a) Extract 323 ($19 \times 17$) $121\times121$-pixel ($6.05\arcsec\times6.05\arcsec$) images and subtract a linear-gradient background for 19 isolated stars in these 17 HST/ACS observations of the Crab; 
(b) register images of stars and use singular value decomposition (SVD) to determine a basis describing the PSF (first SVD component shown); and 
(c) model the pulsar image using the first 72 SVD components.
\label{f:a:1}}
\end{center}
\end{figure}

\begin{figure}[ht]
\begin{center}
\includegraphics[angle=0,width=0.9\columnwidth]{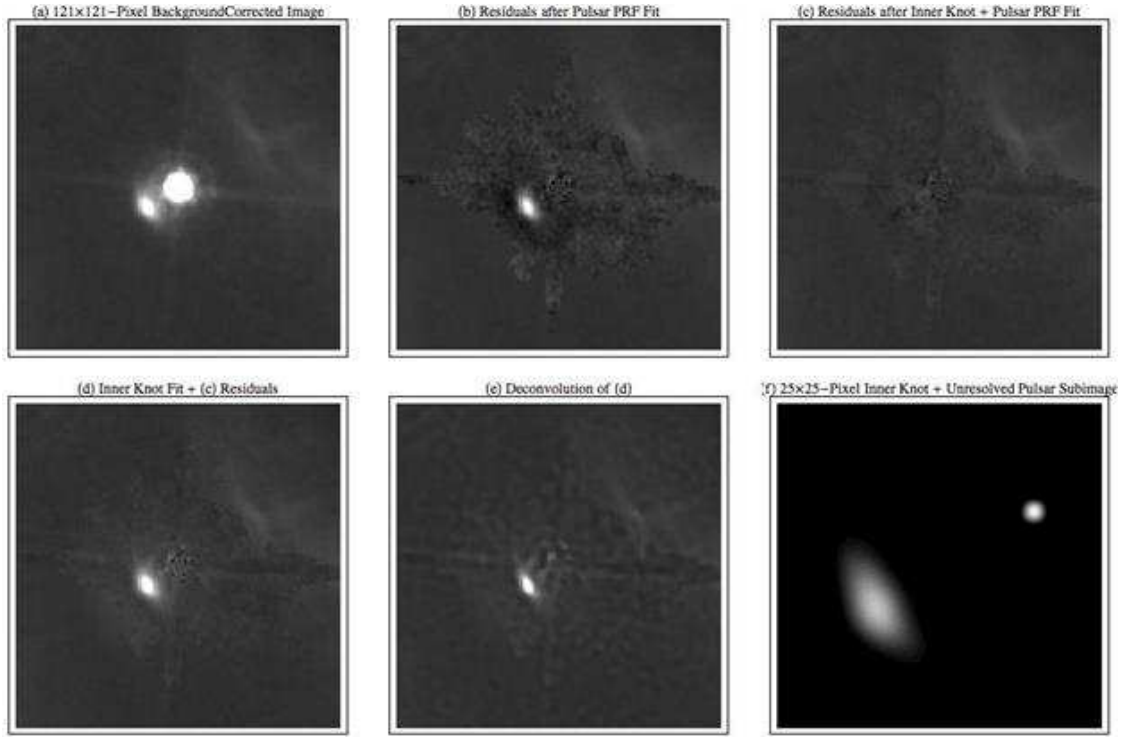}
\figcaption{Images further illustrating steps in processing HST/ACS images of the Crab pulsar and inner knot: 
(a) Subtract a linear-gradient background in central $121\times121$ pixel ($6.05\arcsec\times6.05\arcsec$) image of pulsar and inner knot; 
(b) remove pulsar using SVD model of PSF leaving inner knot; 
(c) remove inner knot using its SVD model leaving residual background; 
(d) add SVD model of inner knot to residual background; 
(e) apply Richardson-Lucy algorithm to generate deconvolved image of inner knot (and residual background); and 
(f) synthesize $25\times25$ pixel ($1.25\arcsec\times1.25\arcsec$) sub-image of registered (unresolved) pulsar and SVD model of inner knot.
\label{f:a:2}}
\end{center}
\end{figure}

In order to remove the pulsar from each of the 17 central $121\times121$-pixel ($6.05\arcsec\times6.05\arcsec$) HST images, we first determine the PSF of the HST/ACS using $121\times121$-pixel images around each of 19 isolated stars in the field. 
In doing this, we account for the fact that some observations were at a roll angle $\sim180\deg$ opposite that of the others. 
After subtracting a fitted linear gradient from the image of each (Figure~\ref{f:a:1}) of the 19 stars and for each of the 17 observations, we register the resulting $323 = 19\times17$ star images and use singular-value decomposition (SVD) to generate a linear basis (lowest term shown in Figure~\ref{f:a:1}) describing the PSF. 
The model pulsar image (Figure~\ref{f:a:1}) uses the first 72 components of the SVD basis. 
After subtracting a linear gradient from each of the 17 central (pulsar) images (Figure~\ref{f:a:2}), we fit each using the PSF basis to generate 17 residual (pulsar-removed) images (also Figure~\ref{f:a:2}), each now dominated by the inner knot. 

{\bf Characterize inner-knot properties.}

\begin{figure}[ht]
\begin{center}s
\includegraphics[angle=0,width=0.9\columnwidth]{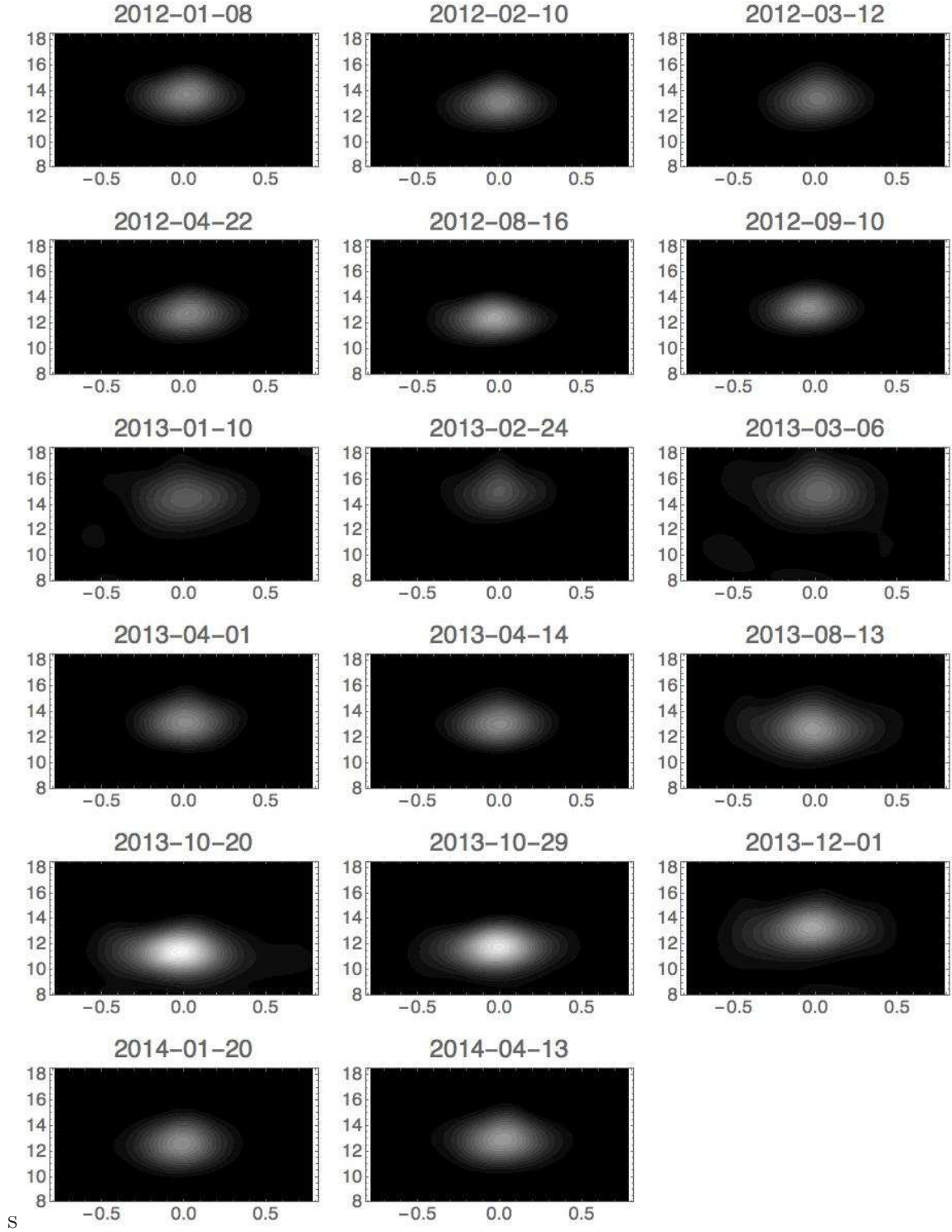}
\figcaption{Mapping of processed images of the inner knot onto a $\psi$--$r$ grid. 
Here $\psi$ is the angular displacement in radian from the knot's centroid; $r$, the radial distance from the pulsar in ($0.05\arcsec$) pixels.
\label{f:a:3}}
\end{center}
\end{figure}

The measured extrinsic properties of the inner knot are the projected radial separation $r_0$ and polar angle $\psi_0$ of the peak surface brightness with respect to the pulsar. 
To facilitate characterization of the intrinsic properties of the inner knot, we map each inner-knot sub-image (Figure~\ref{f:a:2}) onto a $\psi$--$r$ grid (Figure~\ref{f:a:3}). 
For an initial model of the surface brightness (intensity) distribution of the inner knot, we use a simple bivariate normal distribution:
$S(r,\psi) = S_k \exp(-\frac{1}{2}((r-r_0)^2/\sigma_r^2+(\psi-\psi_0)^2/\sigma_\psi^2))$.

The major intrinsic properties of the inner knot are its radial dispersion $\sigma_r$ (FWHM$_r = 2.35 \sigma_r$ for a Gaussian profile), azimuthal dispersion $\sigma_\psi$ (or tangential dispersion $\sigma_t = r_0\sigma_\psi$), and peak intensity (surface brightness) $S_k$. 
Integrating the intensity over solid angle gives the flux $F_k = 2 \pi \sigma_r \sigma_t S_k$. 
In fitting this model to the $\psi$--$r$ image of the inner knot, we include a constant-surface-brightness background $S_b$ as a model parameter.

While this simple model provides adequate estimates to characterize the primary properties of the knot (Table ~\ref{t:hstsvdprop}), as a check we also introduced a somewhat more complicated model to deal with minor asymmetries in the $\psi$--$r$ plane. 
In particular, we took the radial location and radial width to be weak functions of the azimuthal angle, which we expand as a Taylor series to second order:
$r\approx r(\psi)\approx r_0+r'_0(\psi-\psi_0)+\frac{1}{2}r''_0(\psi-\psi_0)^2$
and
$\sigma_r\approx \sigma_r(\psi) \approx \sigma_{r_0}+\sigma'_{r_0}(\psi- \psi_0)+\frac{1}{2}\sigma''_{r_0}(\psi-\psi_0)^2$.
However, the derivatives proved not to alter significantly the results.

{\bf Comparison between methods}

\begin{figure}[ht]
\begin{center}
\includegraphics[angle=0,width=1\columnwidth]{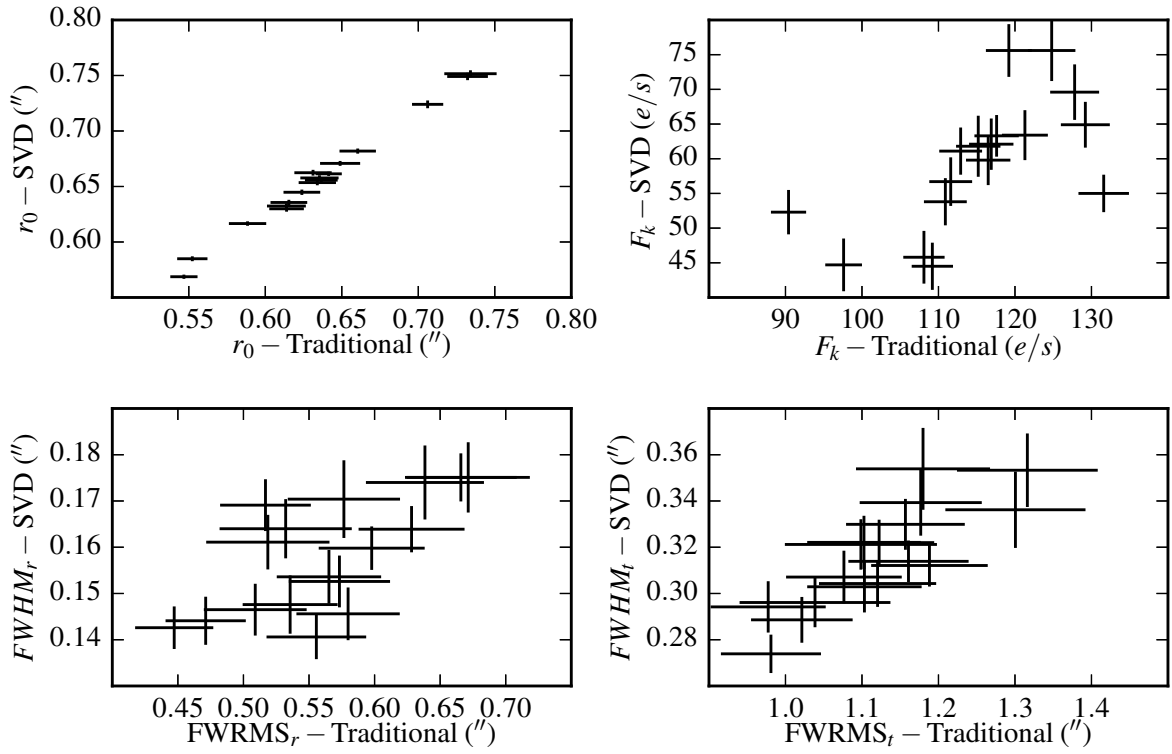}
\figcaption{Comparison of HST-SVD and HST-traditional measured properties of the inner knot.
\label{f:compare}}
\end{center}
\end{figure}

The comparison between methods and the reasons for somewhat emphasizing the SVD-based results considers the various measured parameters individually.
Figure~\ref{f:compare} compares measurements of similar variables between the two methods and illustrates the fact that the two methods.
From the figure we see that, with the possible exception of the knot flux, both methods yield measurements that track each other apart from a scale factor. 
This leads to the result that the pulsar--knot separation determines all the intrinsic properties of the knot. 

Differences for $r_0$ are quite small, whereas the knot flux and size estimates differ by factors of $\approx 2$. 
We would expect differences in the size estimates as the two analyses are using different measures of knot size---the first, RMS full width (FWRMS); the second, FWHM---which are not simply related for a non-Gaussian profile.
That the fluxes differ is not perhaps so surprising as its measurement is sensitive to the method of background subtraction and the impact of emission in the wings of the inner knot's profile.
A significant difference between the two methods is the smaller uncertainty in $r_0$ afforded by the SVD approach, which removes those SVD components that represent noise, thus enhancing signal to noise.

\subsection{{\sl Chandra}}\label{ss:da:chandra}

\begin{figure}[ht]
\begin{center}
\includegraphics[angle=-90,width=0.9\columnwidth]{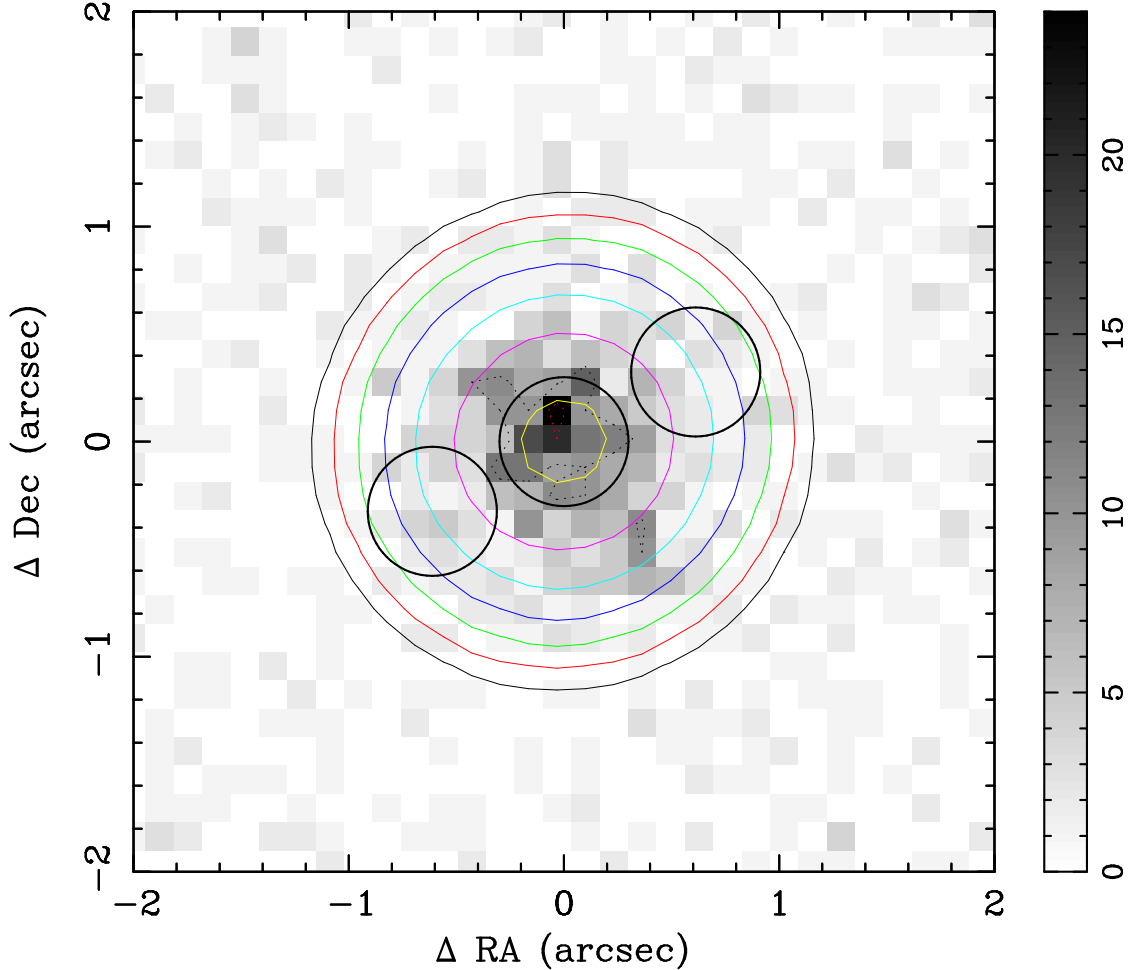}
\figcaption{{\sl Chandra} image at pulse minimum from the sum of the last 6 flaring data sets in Table \ref{t:chandraobs}. 
The 3 small black circles show $0.3\arcsec$-radius extraction regions, the central of which is used to estimate the pulsar flux. 
The circle, at a position angle $\psi = 120\degree$ east of north and $0.65\arcsec$ from the pulsar, is roughly centered on the average optical location of the knot. 
Data in the opposite small circle (the one to the NW where N is up) are used to estimate background. 
Large circles show the best-fit Gaussian to the phase-average pulsar data at intensity levels of 10, 20, 40, 80, 160, 320 and 640 cts/pixel illustrating the level of impact of the PSF at the site of the knot---i.e., slight but non-negligible. 
The intensity level is shown by the grayscale bar on the right.
\label{f:chandra1}}
\end{center}
\end{figure}

Figure~\ref{f:chandra1} shows the summed image at pulse minimum from the 6 ``flaring'' observations (ObsIDs 14684-16247 in Table~\ref{t:chandraobs}). 
Prior to the binning used to produce the figure, we extract the counts in the 3 regions shown. 
It is difficult, if not impossible, to make use of the absolute number of counts as the different instrument configurations are all uncalibrated. 
However, we can make use of the relative numbers of counts from the different regions. 
Thus, the central region serves as an indicator of the number of pulsar counts which can then be compared with the other observations to derive physical fluxes (assuming of course that the phase averaged pulsar flux has not varied). 
The region to the southeast provides the upper limit for the flux in the knot and we use the region to the  northwest to estimate the background, comprised mainly of that due to the wings of the pulsar PSF.

The alert reader will notice a slight excess of counts roughly $0.6\arcsec$ to the southwest of the pulsar. 
This feature is seen both in the phase-averaged pulse and in the data from ObsID 11245. 
(The asymmetric effect of the blade inserted for Obs ID 9765 does not allow these data to be used in making this comparison.) 
Therefore, we conclude that this feature is part of the pulsar PSF. 
A more subtle effect is also present as there is a slight asymmetry in the PSF that causes the region we associate with the inner knot to be somewhat brighter than the background during observations taken in the fall. 
In the spring, when the spacecraft has (naturally) been set at a roll angle that differs by $\approx180\deg$ from that in the fall, the background region is brighter than that we associate with the location of the inner knot. 
This effect is mainly seen in the phase averaged data, and by comparing data from the ObsID 9765 (spring) and ObsID 11245 (fall) observations. 
Both effects described in this paragraph are relatively minor, are mainly due to the fact we are working slightly below the spatial resolution of {\sl Chandra}, and have thus been ignored in setting the upper limits in Table~\ref{t:chandraprop}

\clearpage

\begin{table}[ht]
\begin{center}
\caption {Summary of the Keck observations \label{t:keckobs}}
\begin{tabular}{ccclc} \\ \hline 
\#  & Date       & MJD$^a$   & Instrument$^b$ & Exposure (s) \\ \hline
K1  & 2012-02-08 & 55965.285 & NIRC2-wide     & 420  \\
K2  & 2012-03-05 & 55991.319 & NIRC2-wide     & 420  \\
K3  & 2012-12-22 & 56283.408 & NIRC2-wide     & 1250 \\
K4  & 2012-12-23 & 56284.410 & NIRC2-narrow   & 870  \\
K5  & 2012-12-24 & 56285.388 & NIRC2-narrow   & 1350 \\
K6  & 2012-12-25 & 56286.390 & NIRC2-wide     & 1125 \\
K7  & 2013-02-06 & 56329.308 & NIRC2-narrow   & 2020 \\
K8  & 2013-10-22 & 56587.448 & NIRC2-narrow   & 640  \\
K9  & 2014-01-09 & 56666.274 & NIRC2-wide     & 1560 \\
K10 & 2014-01-09 & 56666.377 & NIRC2-narrow   & 2580 \\
K11 & 2014-01-17 & 56674.325 & NIRC2-wide     & 2270 \\
K12 & 2014-01-17 & 56674.394 & NIRC2-narrow   & 1260 \\
\hline
\end{tabular}
\end{center}
$^a$ Quoted MJD is at the midpoint of the observation.\\
$^b$ Plate scale is 0.04$\arcsec$/pixel for NIRC2-wide; 0.01$\arcsec$/pixel for NIRC2-narrow. 
\end{table}

\begin{table}[ht]
\begin{center}
\caption {Summary of the HST observations \label{t:hstobs} }
\begin{tabular}{ccc} \\ \hline 
\#  & Date       & MJD$^a$$^b$\\ \hline
H1  & 2012-01-08 & 55934.797 \\
H2  & 2012-02-10 & 55967.170 \\
H3  & 2012-03-12 & 55998.414 \\
H4  & 2012-04-22 & 56039.850 \\
H5  & 2012-08-16 & 56155.336 \\
H6  & 2012-09-10 & 56180.960 \\
H7  & 2013-01-10 & 56302.992 \\
H8  & 2013-02-24 & 56347.055 \\
H9  & 2013-03-06 & 56357.023 \\
H10 & 2013-04-01 & 56383.320 \\
H11 & 2013-04-14 & 56396.113 \\
H12 & 2013-08-13 & 56517.710 \\
H13 & 2013-10-20 & 56585.793 \\
H14 & 2013-10-29 & 56594.700 \\
H15 & 2013-12-01 & 56627.273 \\ 
H16 & 2014-01-20 & 56677.530 \\
H17 & 2014-04-13 & 56760.900 \\
\hline
\end{tabular}
\end{center}
$^a$ Quoted MJD is at the midpoint of the observation.\\
$^b$ Plate scale is 0.05$\arcsec$/pixel.
\end{table}

\begin{table}[ht]
\begin{center}
\caption {Summary of the {\sl Chandra} observations \label{t:chandraobs} }
\begin{tabular}{ccccc} \\ \hline 
\#  & ObsID & Date & MJD$^a$ & Exposure (s)\\ \hline
C1  & 09765 & 2008-01-22  & 54487.670 & 95082 \\
C2  & 11245 & 2010-11-16  & 55516.294 & 22039 \\
C3  & 14684 & 2013-03-05  & 56356.972 & 19485 \\
C4  & 14686 & 2013-03-10  & 56361.338 & 20052 \\
C5  & 14687 & 2013-03-17  & 56368.324 & 16718 \\
C6  & 16244 & 2013-10-19  & 56584.658 & 20048 \\
C7  & 16246 & 2013-10-22  & 56587.515 & 20013 \\
C8  & 16247 & 2013-10-24  & 56589.007 & 19770 \\
\hline
\end{tabular}
\end{center}
$^a$ Quoted MJD is at the start of the observation.
\end{table}

\clearpage

\begin{table}[ht]
\begin{center}
\caption{Knot properties based upon analysis of the Keck data \label{t:keckprop}}
\begin{tabular}{cccccc} \\ \hline
    &  $\psi_0$         & $r_0$               & FWHM$_r$            & FWHM$_t$            & $F_k$               \\
\#  &  $\degree$        & $\arcsec$           & $\arcsec$           & $\arcsec$           & $F_p$               \\ \hline
K1  & $118.50 \pm 0.16$ & $0.6711 \pm 0.0015$ & $0.3931 \pm 0.0040$ & $0.4719 \pm 0.0051$ & $0.0639 \pm 0.0045$ \\
K2  & $121.00 \pm 0.13$ & $0.6577 \pm 0.0015$ & $0.4121 \pm 0.0035$ & $0.4252 \pm 0.0042$ & $0.0626 \pm 0.0067$ \\
K3  & $122.64 \pm 0.11$ & $0.6893 \pm 0.0011$ & $0.3215 \pm 0.0028$ & $0.3948 \pm 0.0039$ & $0.0804 \pm 0.0074$ \\
K4  & $120.96 \pm 0.18$ & $0.7118 \pm 0.0021$ & $0.3401 \pm 0.0051$ & $0.4081 \pm 0.0059$ & $0.0473 \pm 0.0003$ \\
K5  & $121.20 \pm 0.07$ & $0.6981 \pm 0.0007$ & $0.2983 \pm 0.0019$ & $0.3767 \pm 0.0024$ & $0.0555 \pm 0.0005$ \\
K6  & $120.79 \pm 0.12$ & $0.7164 \pm 0.0010$ & $0.3375 \pm 0.0029$ & $0.4238 \pm 0.0042$ & $0.0599 \pm 0.0085$ \\
K7  & $121.93 \pm 0.12$ & $0.7756 \pm 0.0013$ & $0.3355 \pm 0.0034$ & $0.4049 \pm 0.0038$ & $0.0662 \pm 0.0004$ \\
K8  & $120.33 \pm 0.06$ & $0.5722 \pm 0.0004$ & $0.2216 \pm 0.0011$ & $0.3172 \pm 0.0015$ & $0.0621 \pm 0.0005$ \\
K9  & $121.46 \pm 0.08$ & $0.6543 \pm 0.0006$ & $0.2958 \pm 0.0018$ & $0.4030 \pm 0.0024$ & $0.0708 \pm 0.0047$ \\
K10 & $115.77 \pm 0.03$ & $0.6257 \pm 0.0003$ & $0.2689 \pm 0.0007$ & $0.3683 \pm 0.0009$ & $0.0663 \pm 0.0006$ \\
K11 & $121.49 \pm 0.10$ & $0.6371 \pm 0.0008$ & $0.3119 \pm 0.0022$ & $0.3787 \pm 0.0029$ & $0.0647 \pm 0.0078$ \\
K12 & $118.99 \pm 0.09$ & $0.6438 \pm 0.0009$ & $0.3248 \pm 0.0023$ & $0.3766 \pm 0.0025$ & $0.0574 \pm 0.0001$ \\ \hline
\end{tabular}
\end{center}
\end{table}

\begin{table}[ht]
\begin{center}
\caption {Results of regression analyses for the 12 Keck observations \label{t:keckregression} }
\begin{tabular}{ccccc} \\ \hline 
Power-law for $r_1 = 0.6692\arcsec$  &                     &                    &             \\
Property    & Unit                   & $p(r_1)$            & Power-law index $q$ & Probability \\ \hline
FWHM$_r$    & $\arcsec$              & $0.3180 \pm 0.0132$ &  $ 1.19 \pm 0.54$  & 5.4E-2      \\
FWHM$_t$    & $\arcsec$              & $0.3941 \pm 0.0093$ &  $ 0.75 \pm 0.31$  & 3.8E-2      \\
$S_k$       & $F_p/(\arcsec)^2$      & $0.439 \pm 0.033$   &  $-2.14 \pm 0.97$  & 5.2E-2      \\
$F_k$       & $F_p$                  & $0.0626 \pm 0.0025$ &  $-0.20 \pm 0.53$  & 7.1E-1      \\ \hline
Linear for $<r_0> = 0.6711\arcsec$   &                     &                    &             \\ 
Property    &                        & $\psi_0(<r_0>)$     &  $d\psi_0/dr_0$    &             \\ \hline
$\psi_0$  & $\degree$, $\degree/\arcsec$ & $120.42 \pm 0.51$ & $15.2 \pm 10.2$  & 1.7E-1   \\ \hline
\end{tabular}
\end{center}
\end{table}

\begin{table}
\begin{center}
\caption{Knot properties based upon SVD analysis of the 17 HST observations (see Appendix) \label{t:hstsvdprop}}
\tiny
\begin{tabular} {ccccccccc} \\ \hline
\#  & $\psi_0$          & $r_0$               & FWHM$_r$            & FWHM$_t$            & $S_k$        & $F_k$          & $F_p$          & $F_k$           \\
    & $\degree$         & $\arcsec$           & $\arcsec$           & $\arcsec$         & e/s/$(\arcsec)^2$ & e/s         & e/s            & $F_p$                      \\ \hline
H1  & $120.54 \pm 0.40$ & $0.6819 \pm 0.0023$ & $0.1526 \pm 0.0056$ & $0.3221 \pm 0.0116$ & $1064 \pm 33$& $59.8 \pm 3.6$ & $1425 \pm 82$  & $0.0420 \pm 0.0035$ \\
H2  & $119.45 \pm 0.41$ & $0.6535 \pm 0.0024$ & $0.1536 \pm 0.0058$ & $0.3071 \pm 0.0114$ & $1008 \pm 32$& $53.8 \pm 3.4$ & $1433 \pm 80$  & $0.0375 \pm 0.0032$ \\
H3  & $120.37 \pm 0.33$ & $0.6708 \pm 0.0022$ & $0.1751 \pm 0.0052$ & $0.3121 \pm 0.0092$ & $ 902 \pm 23$& $55.0 \pm 2.6$ & $1471 \pm 64$  & $0.0374 \pm 0.0024$ \\
H4  & $121.74 \pm 0.41$ & $0.6356 \pm 0.0024$ & $0.1465 \pm 0.0056$ & $0.2942 \pm 0.0110$ & $1053 \pm 34$& $52.3 \pm 3.2$ & $1276 \pm 80$  & $0.0410 \pm 0.0035$ \\
H5  & $117.90 \pm 0.39$ & $0.6166 \pm 0.0020$ & $0.1406 \pm 0.0048$ & $0.3043 \pm 0.0102$ & $1256 \pm 36$& $61.1 \pm 3.4$ & $1231 \pm 84$  & $0.0497 \pm 0.0044$ \\
H6  & $117.62 \pm 0.41$ & $0.6577 \pm 0.0024$ & $0.1456 \pm 0.0058$ & $0.3029 \pm 0.0112$ & $1102 \pm 36$& $56.7 \pm 3.5$ & $1184 \pm 86$  & $0.0479 \pm 0.0046$ \\
H7  & $121.17 \pm 0.59$ & $0.7240 \pm 0.0035$ & $0.1704 \pm 0.0084$ & $0.3539 \pm 0.0176$ & $ 675 \pm 28$& $44.7 \pm 3.9$ & $1422 \pm 77$  & $0.0314 \pm 0.0032$ \\
H8  & $119.91 \pm 0.50$ & $0.7516 \pm 0.0031$ & $0.1751 \pm 0.0076$ & $0.3533 \pm 0.0158$ & $ 667 \pm 25$& $44.5 \pm 3.4$ & $1421 \pm 65$  & $0.0313 \pm 0.0028$ \\
H9  & $120.88 \pm 0.52$ & $0.7491 \pm 0.0033$ & $0.1740 \pm 0.0080$ & $0.3362 \pm 0.0166$ & $ 728 \pm 29$& $45.8 \pm 3.8$ & $1359 \pm 73$  & $0.0337 \pm 0.0033$ \\
H10 & $120.60 \pm 0.32$ & $0.6614 \pm 0.0019$ & $0.1598 \pm 0.0046$ & $0.3139 \pm 0.0090$ & $1098 \pm 27$& $63.3 \pm 3.0$ & $1295 \pm 72$  & $0.0489 \pm 0.0036$ \\
H11 & $119.39 \pm 0.36$ & $0.6557 \pm 0.0021$ & $0.1639 \pm 0.0050$ & $0.3220 \pm 0.0098$ & $1090 \pm 28$& $64.9 \pm 3.3$ & $1509 \pm 77$  & $0.0430 \pm 0.0031$ \\
H12 & $119.28 \pm 0.41$ & $0.6324 \pm 0.0023$ & $0.1691 \pm 0.0056$ & $0.3299 \pm 0.0110$ & $1028 \pm 29$& $63.4 \pm 3.6$ & $1283 \pm 83$  & $0.0494 \pm 0.0042$ \\
H13 & $117.85 \pm 0.36$ & $0.5687 \pm 0.0019$ & $0.1426 \pm 0.0046$ & $0.2739 \pm 0.0082$ & $1668 \pm 44$& $75.6 \pm 3.8$ & $1222 \pm 106$ & $0.0618 \pm 0.0062$ \\
H14 & $118.82 \pm 0.41$ & $0.5849 \pm 0.0022$ & $0.1441 \pm 0.0052$ & $0.2886 \pm 0.0098$ & $1583 \pm 48$& $75.6 \pm 4.4$ & $1285 \pm 118$ & $0.0589 \pm 0.0064$ \\
H15 & $118.59 \pm 0.52$ & $0.6624 \pm 0.0026$ & $0.1476 \pm 0.0061$ & $0.3393 \pm 0.0144$ & $1105 \pm 40$& $61.8 \pm 4.4$ & $1353 \pm 100$ & $0.0457 \pm 0.0047$ \\
H16 & $118.82 \pm 0.41$ & $0.6299 \pm 0.0026$ & $0.1640 \pm 0.0064$ & $0.2961 \pm 0.0106$ & $1084 \pm 35$& $62.1 \pm 3.7$ & $1419 \pm 98$  & $0.0438 \pm 0.0040$ \\
H17 & $120.88 \pm 0.41$ & $0.6448 \pm 0.0024$ & $0.1611 \pm 0.0058$ & $0.3212 \pm 0.0110$ & $1164 \pm 35$& $69.6 \pm 4.0$ & $1177 \pm 99$  & $0.0591 \pm 0.0060$ \\
 \hline
\end{tabular}
\end{center}
\end{table}

\begin{table}[ht]
\begin{center}
\caption {Results of regression analyses for the 17 HST observations based on the data in Table~\ref{t:hstsvdprop}  \label{t:hstregression} }
\begin{tabular}{ccccc} \\ \hline 
Power-law  for $r_1 = 0.6559\arcsec$  &                     &                   &             \\
Property   & Unit                     & $p(r_1)$          & Power-law index $q$ & Probability \\ \hline
FWHM$_r$   & $\arcsec$                & $0.1575 \pm 0.0022$ & $0.72 \pm 0.19$   & 1.7E-3      \\
FWHM$_t$   & $\arcsec$                & $0.3152 \pm 0.0029$ & $0.80 \pm 0.13$   & 1.4E-5      \\
$S_k$      & $F_p/(\arcsec)^2$        & $0.782 \pm 0.023$   & $-3.66 \pm 0.40$  & 1.7E-7      \\
$F_k$      & $F_p$                    & $0.0440 \pm 0.0012$ & $-2.39 \pm 0.37$  & 9.8E-6      \\ \hline
Linear   for $<r_0> = 0.6577\arcsec$  &                     &                   &             \\
Property   &                          & $\psi_0(<r_0> )$    & $d\psi_0/dr_0$    &             \\ \hline
$\psi_0$   & $\degree$, $\degree/\arcsec$ & $119.64 \pm 0.27$ & $13.5 \pm 5.5$  & 2.7E-2      \\ \hline
\end{tabular}
\end{center}
\end{table}

\begin{table}[ht]
\begin{center}
\caption {Results of the {\sl Chandra} analysis \label{t:chandraprop} }
\begin{tabular}{cccccccc} \\ \hline 
 & \multicolumn{3}{c}{Pulse Average} & \multicolumn{3}{c}{Pulse Minimum} & $F_k/F_p$ \\ \hline
ObsID  & Pulsar  & Knot & Background & Pulsar & Knot & Background & $3\sigma$ \\ \hline
09765  & 51844   & 2963  & 3704  & 845  & 57  &  72  & $<0.0022$ \\
11245  &  5630   & 1018  &  851  &  88  & 16  &  12  & $<0.0094$ \\
last 6 & 13549   & 2622  & 2503  & 203  & 39  &  44  & $<0.0067$ \\ \hline
\end{tabular}
\end{center}
\end{table}

\begin{table}[ht]
\begin{center}
\caption {{\sl Fermi}/LAT 12-hr-average fluxes and pulsar--knot separations as measured with Keck and HST (SVD analysis) 
\label{t:fermi} }
\begin{tabular}{lllll} \\ \hline 
\#  & Date       & $t_{\rm mid}$ & $r_0$  & {\sl Fermi}/LAT $\geq 100$ MeV \\
    &            & MJD       & $\arcsec$  & $10^{-6}$ ph/s/cm$^2$ \\ \hline    
H1  & 2012-01-08 & 55934.797 & $0.6819 \pm 0.0023$ &  2.45 (+0.47, -0.43) \\
K1  & 2012-02-08 & 55965.280 & $0.6711 \pm 0.0015$ &  2.58 (+0.60, -0.54) \\
H2  & 2012-02-10 & 55967.171 & $0.6535 \pm 0.0024$ &  2.79 (+0.65, -0.58) \\
K2  & 2012-03-05 & 55991.311 & $0.6577 \pm 0.0015$ &  2.78 (+0.46, -0.43) \\
H3  & 2012-03-12 & 55998.414 & $0.6708 \pm 0.0022$ &  3.55 (+0.52, -0.49) \\
H4  & 2012-04-22 & 56039.850 & $0.6356 \pm 0.0024$ &  2.18 (+0.56, -0.50) \\
H5  & 2012-08-16 & 56155.336 & $0.6166 \pm 0.0020$ &  3.25 (+0.56, -0.52) \\
H6  & 2012-09-10 & 56180.960 & $0.6577 \pm 0.0024$ &  3.86 (+0.50, -0.47) \\
K3  & 2012-12-22 & 56283.381 & $0.6893 \pm 0.0011$ &  3.13 (+0.68, -0.63) \\
K4  & 2012-12-23 & 56284.403 & $0.7118 \pm 0.0021$ &  3.51 (+0.67, -0.62) \\
K5  & 2012-12-24 & 56285.379 & $0.6981 \pm 0.0007$ &  3.29 (+0.67, -0.62) \\
K6  & 2012-12-25 & 56286.370 & $0.7164 \pm 0.0010$ &  2.74 (+0.62, -0.56) \\
H7  & 2013-01-10 & 56302.992 & $0.7240 \pm 0.0035$ &  3.22 (+0.66, -0.66) \\
K7  & 2013-02-06 & 56329.248 & $0.7756 \pm 0.0013$ &  3.93 (+0.75, -0.70) \\
H8  & 2013-02-24 & 56347.055 & $0.7516 \pm 0.0031$ &  5.44 (+0.72, -0.68) \\
H9  & 2013-03-06 & 56357.023 & $0.7491 \pm 0.0033$ & 10.09 (+0.52, -0.51) \\
H10 & 2013-04-01 & 56383.320 & $0.6614 \pm 0.0019$ &  3.50 (+0.68, -0.63) \\
H11 & 2013-04-14 & 56396.113 & $0.6557 \pm 0.0021$ &  2.89 (+0.59, -0.54) \\
H12 & 2013-08-13 & 56517.710 & $0.6324 \pm 0.0023$ &  2.65 (+0.46, -0.43) \\
H13 & 2013-10-20 & 56585.793 & $0.5687 \pm 0.0019$ &  3.71 (+0.26, -0.25) \\
K8  & 2013-10-22 & 56587.443 & $0.5722 \pm 0.0004$ &  1.25 (+1.73, -1.73) \\
H14 & 2013-10-29 & 56594.700 & $0.5849 \pm 0.0022$ &  7.46 (+1.16, -1.07) \\
H15 & 2013-12-01 & 56627.273 & $0.6624 \pm 0.0026$ &  2.15 (+0.53, -0.47) \\
K9  & 2014-01-09 & 56666.316 & $0.6543 \pm 0.0006$ &  3.38 (+0.53, -0.50) \\
K10 & 2014-01-09 & 56666.438 & $0.6257 \pm 0.0003$ &  3.54 (+0.51, -0.48) \\ 
K11 & 2014-01-17 & 56674.356 & $0.6371 \pm 0.0008$ &  3.84 (+0.53, -0.50) \\
K12 & 2014-01-17 & 56674.400 & $0.6438 \pm 0.0009$ &  3.47 (+0.44, -0.42) \\   
H16 & 2014-01-20 & 56677.530 & $0.6299 \pm 0.0026$ &  2.85 (+0.43, -0.41) \\
H17 & 2014-04-13 & 56760.900 & $0.6448 \pm 0.0024$ &  3.96 (+0.78, -0.71) \\ \hline
\end{tabular}
\end{center}
\end{table}

\begin{table} [ht]
\begin{center}
\caption{Results of the Traditional HST Analysis \label{t:hsttrad}}
\begin{tabular}{ccccccccc} \\ \hline
\#  & $r_0$             & FWRMS$_r$         & FWRMS$_t$         & $F_k$           \\ 
    & $\arcsec$         & $\arcsec$         & $\arcsec$         & e/s             \\ \hline
H1  & $0.660 \pm 0.012$ & $0.287 \pm 0.019$ & $0.551 \pm 0.037$ & $116.5 \pm 2.9$ \\
H2  & $0.634 \pm 0.012$ & $0.283 \pm 0.020$ & $0.538 \pm 0.038$ & $110.9 \pm 2.8$ \\
H3  & $0.649 \pm 0.013$ & $0.333 \pm 0.021$ & $0.594 \pm 0.038$ & $131.6 \pm 3.3$ \\
H4  & $0.615 \pm 0.012$ & $0.255 \pm 0.020$ & $0.489 \pm 0.038$ & $ 90.4 \pm 2.3$ \\
H5  & $0.588 \pm 0.012$ & $0.278 \pm 0.019$ & $0.560 \pm 0.038$ & $112.9 \pm 2.8$ \\
H6  & $0.635 \pm 0.012$ & $0.290 \pm 0.020$ & $0.552 \pm 0.037$ & $111.6 \pm 2.8$ \\
H7  & $0.706 \pm 0.010$ & $0.288 \pm 0.021$ & $0.589 \pm 0.044$ & $ 97.6 \pm 2.4$ \\
H8  & $0.734 \pm 0.017$ & $0.336 \pm 0.024$ & $0.658 \pm 0.046$ & $109.2 \pm 2.7$ \\
H9  & $0.732 \pm 0.013$ & $0.319 \pm 0.022$ & $0.650 \pm 0.046$ & $108.1 \pm 2.7$ \\
H10 & $0.641 \pm 0.009$ & $0.299 \pm 0.020$ & $0.580 \pm 0.039$ & $117.6 \pm 2.9$ \\
H11 & $0.637 \pm 0.011$ & $0.314 \pm 0.020$ & $0.561 \pm 0.036$ & $129.2 \pm 3.2$ \\
H12 & $0.613 \pm 0.012$ & $0.258 \pm 0.017$ & $0.578 \pm 0.039$ & $121.3 \pm 3.0$ \\
H13 & $0.547 \pm 0.009$ & $0.224 \pm 0.015$ & $0.491 \pm 0.033$ & $119.2 \pm 3.0$ \\
H14 & $0.552 \pm 0.009$ & $0.236 \pm 0.015$ & $0.511 \pm 0.033$ & $124.8 \pm 3.1$ \\
H15 & $0.631 \pm 0.012$ & $0.268 \pm 0.018$ & $0.588 \pm 0.040$ & $115.2 \pm 2.9$ \\
H16 & $0.614 \pm 0.011$ & $0.266 \pm 0.025$ & $0.519 \pm 0.049$ & $116.9 \pm 2.9$ \\
H17 & $0.624 \pm 0.012$ & $0.259 \pm 0.024$ & $0.549 \pm 0.050$ & $127.8 \pm 3.2$ \\ \hline
\end{tabular}
\end{center}
\end{table}

\begin{table}[ht]
\begin{center}
\caption {Results of regression analyses for the 17 HST observations based on the traditional data analysis \label{t:hsttradregression} }
\begin{tabular}{ccccc} \\ \hline 
Power-law for $r_1 = 0.6342\arcsec$  &                      &                   &             \\
Property   & Unit                    & $p(r_1)$           & Power-law index $q$ & Probability \\ \hline
FWRMS$_r$  & $\arcsec$               & $0.2801 \pm 0.0046$  & $1.14 \pm 0.21$   & 6.0E-5      \\ 
FWRMS$_t$  & $\arcsec$               & $0.5605 \pm 0.0067$  & $0.85 \pm 0.15$   & 4.9E-5      \\
$F_k$      & e/s                     & $114.8 \pm 2.6$      & $-0.39 \pm 0.29$  & 1.9E-1      \\ \hline
\end{tabular}
\end{center}
\end{table}

\clearpage


\begin{thebibliography}{}

\bibitem[H.~E.~S.~S.~Collaboration et al.(2014)]{Abra14} H.~E.~S.~S.~Collaboration, Abramowski, A., Aharonian, F., et al.\ 2014, \aap, 562, LL4 

\bibitem[Aliu et al.(2014)]{Aliu14} Aliu, E., Archambault, S., Aune, T., et al.\ 2014, \apjl, 781, LL11 

\bibitem[{{Arons}(2012)}]{Arons2012SSRv..173..341A} {Arons}, J. 2012, \ssr, 173, 341

\bibitem[{{Baty} {et~al.}(2013){Baty}, {Petri}, \&
  {Zenitani}}]{Baty2013MNRAS.436L..20B}
{Baty}, H., {Petri}, J., \& {Zenitani}, S. 2013, \mnras, 436, L20

\bibitem[Beckwith et al.(2006)]{Beck06} Beckwith, S.~V.~W., Stiavelli, M., Koekemoer, A.~M., et al.\ 2006, \aj, 132, 1729 

\bibitem[Bertin \& Arnouts(1996)]{Bert96} Bertin, E., \& Arnouts, S.\ 1996, \aaps, 117, 393 

\bibitem[{{Bogovalov}(1999)}]{Bogovalov1999A&A...349.1017B} {Bogovalov}, S.~V. 1999, \aap, 349, 1017

\bibitem[B{\"u}hler \& Blandford(2014)]{Bueh14} B{\"u}hler, R., \& Blandford, R.\ 2014, Reports on Progress in Physics, 77, 066901

\bibitem[Buehler et al.(2012)]{Bueh12} Buehler, R., Scargle, J.~D., Blandford, R.~D., et al.\ 2012, \apj, 749, 26 

\bibitem[Buson et al.(2013)]{Buso13} Buson, S., Buehler, R., \& Hays, E.\ 2013, The Astronomer's Telegram, 5485, 1 

\bibitem[{{Bykov} {et~al.}(2012){Bykov}, {Pavlov}, {Artemyev}, \& {Uvarov}}]{Bykov2012MNRAS.421L..67B} {Bykov}, A.~M., {Pavlov}, G.~G., {Artemyev}, A.~V., \& {Uvarov}, Y.~A. 2012, \mnras, 421, L67

\bibitem[Camus et al.(2009)]{Camus09} Camus, N.~F., Komissarov, S.~S., Bucciantini, N., \& Hughes, P.~A.\ 2009, \mnras, 400, 1241 

\bibitem[{{Cerutti} {et~al.}(2012){Cerutti}, {Uzdensky}, \&  {Begelman}}]{Cerutti2012ApJ...746..148C} {Cerutti}, B., {Uzdensky}, D.~A., \& {Begelman}, M.~C. 2012, \apj, 746, 148

\bibitem[{{Cerutti} {et~al.}(2013){Cerutti}, {Werner}, {Uzdensky}, \&  {Begelman}}]{Cerutti2013ApJ...770..147C} {Cerutti}, B., {Werner}, G.~R., {Uzdensky}, D.~A., \& {Begelman}, M.~C. 2013, \apj, 770, 147

\bibitem[{{Clausen-Brown} \& {Lyutikov}(2012)}]{Clausen-Brown2012MNRAS.426.1374C} {Clausen-Brown}, E., \& {Lyutikov}, M. 2012, \mnras, 426, 1374

\bibitem[{{Fitzpatrick}(1999)}]{Fitzpatrick1999PASP..111...63F} {Fitzpatrick}, E.~L. 1999, \pasp, 111, 63

\bibitem[Fruchter \& Hook(2002)]{Fruc02} Fruchter, A.~S., \& Hook, R.~N.\ 2002, \pasp, 114, 144 

\bibitem[Ghez et al.(2008)]{Ghez08} Ghez, A.~M., Salim, S., Weinberg, N.~N., et al.\ 2008, \apj, 689, 1044 

\bibitem[Hester(2008)]{Hest08} Hester, J.~J.\ 2008, \araa, 46, 127 

\bibitem[Hester et al.(1995)]{Hest95} Hester, J.~J., Scowen, P.~A., Sankrit, R., et al.\ 1995, \apj, 448, 240 

\bibitem[Komissarov \& Lyubarsky(2003)]{Komi03} Komissarov, S.~S., \& Lyubarsky, Y.~E.\ 2003, \mnras, 344, L93 

\bibitem[{{Komissarov} \& {Lyutikov}(2011)}]{KomissarovLyutikov2011MNRAS.414.2017K} {Komissarov}, S.~S., \& {Lyutikov}, M. 2011, \mnras, 414, 2017

\bibitem[Kron(1980)]{Kron80} Kron, R.~G.\ 1980, \apjs, 43, 305 

\bibitem[{{Lind} \& {Blandford}(1985)}]{LindBlandford1985ApJ...295..358L} {Lind}, K.~R., \& {Blandford}, R.~D. 1985, \apj, 295, 358

\bibitem[{{Lyubarsky}(2012)}]{Lyubarsky2012MNRAS.427.1497L} {Lyubarsky}, Y.~E. 2012, \mnras, 427, 1497

\bibitem[{{Lyutikov} {et~al.}(2012){Lyutikov}, {Balsara}, \& {Matthews}}]{Lyutikov2012MNRAS.422.3118L} {Lyutikov}, M., {Balsara}, D., \& {Matthews}, C. 2012, \mnras, 422, 3118

\bibitem[{{Lyutikov} {et~al.}(2003){Lyutikov}, {Pariev}, \& {Blandford}}]{Lyutikov2003ApJ...597..998L} {Lyutikov}, M., {Pariev}, V.~I., \& {Blandford}, R.~D. 2003, \apj, 597, 998

\bibitem[Mayer et al.(2013)]{Maye13} Mayer, M., Buehler, R., Hays, E., et al.\ 2013, \apjl, 775, L37 

\bibitem[Melatos et~al.(2005)] {Melatos05}  Melatos, A., Scheltus, D., Whiting, M. T., Eikenberry, S. S., Romani, R. W., Rigaut, F., Spitkovsky, A., Arons, J. \& Payne, D. J. B. 2005, \apj, 633, 931

\bibitem[{{Moran} {et~al.}(2013){Moran}, {Shearer}, {Mignani}, {S{\l}owikowska}, {De Luca}, {Gouiff{\`e}s}, \& {Laurent}}]{Moran2013MNRAS.433.2564M} {Moran}, P., {Shearer}, A., {Mignani}, R.~P., {S{\l}owikowska}, A., {De Luca}, A., {Gouiff{\`e}s}, C., \& {Laurent}, P. 2013, \mnras, 433, 2564

\bibitem[Ng \& Romani (2008)]{Ng08} Ng, C.-Y. \& Romani, R. W. 2008, \apj,673, 411

\bibitem[Ojha et al.(2013)]{Ojha13} Ojha, R., Hays, E., Buehler, R., \& Dutka, M.\ 2013, The Astronomer's Telegram, 4855, 1 

\bibitem[{{Porth} {et~al.}(2014){Porth}, {Komissarov}, \& {Keppens}}]{Porth14} {Porth}, O., {Komissarov}, S.~S., \& {Keppens}, R. 2014, \mnras, 438, 278

\bibitem[{{Rees} \& {Gunn}(1974)}]{ReesGunn1974MNRAS.167....1R} {Rees}, M.~J., \& {Gunn}, J.~E. 1974, \mnras, 167, 1

\bibitem[{{Sandberg} \& {Sollerman}(2009)}]{SandbergSollerman2009A&A...504..525S} {Sandberg}, A., \& {Sollerman}, J. 2009, \aap, 504, 525

\bibitem[{{Sollerman} {et~al.}(2000){Sollerman}, {Lundqvist}, {Lindler}, {Chevalier}, {Fransson}, {Gull}, {Pun}, \& {Sonneborn}}]{Sollerman2000ApJ...537..861S} {Sollerman}, J., {Lundqvist}, P., {Lindler}, D., {Chevalier}, R.~A., {Fransson}, C., {Gull}, T.~R., {Pun}, C.~S.~J., \& {Sonneborn}, G. 2000, \apj, 537, 861

\bibitem[Sollerman(2003)]{Soll03} Sollerman, J.\ 2003, \aap, 406, 639 

\bibitem[Spitkovsky(2006)]{Spit06} Spitkovsky, A.\ 2006, \apjl, 648, L51 

\bibitem[Striani et al.(2011)]{Stri11} Striani, E., Tavani, M., Piano, G., et al.\ 2011, \apjl, 741, L5 

\bibitem[{{Sturrock} \& {Aschwanden}(2012)}]{Sturrock2012ApJ...751L..32S} {Sturrock}, P., \& {Aschwanden}, M.~J. 2012, \apjl, 751, L32

\bibitem[Tavani et al.(2011)]{Tava11} Tavani, M., Bulgarelli, A., Vittorini, V., et al.\ 2011, Science, 331, 736

\bibitem[Tennant et al.(2001)]{Tenn01} Tennant, A.~F., Becker, W., Juda, M., et al.\ 2001, \apjl, 554, L173 

\bibitem[{{Tchekhovskoy} {et~al.}(2013){Tchekhovskoy}, {Spitkovsky}, \& {Li}}]{Tchekhovskoy2013MNRAS.435L...1T} {Tchekhovskoy}, A., {Spitkovsky}, A., \& {Li}, J.~G. 2013, \mnras, 435, L1

\bibitem[Tchekhovskoy, Philippov, \& Spitkovsky (2015)]{Tchek15} Tchekhovskoy, A., Philippov, A. \& Spitkovsky,A. 2015, eprint arXiv:1503.01467

\bibitem[{{Teraki} \& {Takahara}(2013)}]{Teraki2013ApJ...763..131T} {Teraki}, Y., \& {Takahara}, F. 2013, \apj, 763, 131

\bibitem[Tziamtzis et al.(2009)]{Tzia09} Tziamtzis, A., Lundqvist, P., \& Djupvik, A.~A.\ 2009, \aap, 508, 221 

\bibitem[{{Uzdensky} {et~al.}(2011){Uzdensky}, {Cerutti}, \& {Begelman}}]{Uzdensky2011ApJ...737L..40U} {Uzdensky}, D.~A., {Cerutti}, B., \& {Begelman}, M.~C. 2011, \apjl, 737, L40

\bibitem[{Wizinowich et al. (2006)}]{2006PASP..118..297W} Wizinowich, P.~L., Le Mignant, D., Bouchez, A.~H., Campbell, R.~D., Chin, J.~C.~Y. et al. 2006, \pasp, 118, 297

\bibitem[Weisskopf et al.(2013)]{Weis13} Weisskopf, M.~C., Tennant, A.~F., Arons, J., et al.\ 2013, \apj, 765, 56 

\bibitem[{{Yuan} {et~al.}(2011){Yuan}, {Yin}, {Wu}, {Bi}, {Liu}, \& {Zhang}}]{Yuan2011ApJ...730L..15Y} {Yuan}, Q., {Yin}, P.-F., {Wu}, X.-F., {Bi}, X.-J., {Liu}, S., \& {Zhang}, B. 2011, \apjl, 730, L15

\bibitem[Yelda et al.(2010)]{Yeld10} Yelda, S., Lu, J.~R., Ghez, A.~M., et al.\ 2010, \apj, 725, 331 

\end{thebibliography}
\end{document}